\documentclass{jov_arxiv}

\usepackage{amsmath,amssymb}
\usepackage{mathrsfs}
\usepackage{mathbbol}
\usepackage{lscape}
\usepackage{afterpage}
\usepackage{graphicx}
\usepackage{hyperref}
\usepackage{wrapfig}
\usepackage[T1]{fontenc}    
\usepackage{hyperref}       
\usepackage{url}            
\usepackage{booktabs}       
\usepackage{amsfonts}       
\usepackage{nicefrac}       
\usepackage{microtype}      
\usepackage{float}
\usepackage{amsmath}
\usepackage{diagbox}
\usepackage{xcolor}
\usepackage{multirow}

\usepackage{array}
\usepackage{rotating}

\newcommand{\blue}[1]{\textcolor[rgb]{0,0,0}{#1}}
\newcommand{\azul}[1]{\textcolor[rgb]{0,0,1}{#1}}
\newcommand{\green}[1]{\textcolor[rgb]{0,0,0}{#1}}

\newlength\savedwidth
\usepackage[all]{xy}
\usepackage{amsmath}
\usepackage{dsfont}
\usepackage{cancel}
\usepackage{lineno}
\usepackage{pbox}
\usepackage{bbm}
\newcommand{\vect}[1]{\boldsymbol{#1}}
\begin{document}
\setpagewiselinenumbers
\modulolinenumbers[1]

\title{Contrast Sensitivity Functions in Autoencoders\vspace{-0.3cm}}

\abstract{
\blue{Three decades ago, Atick et al. suggested that human frequency sensitivity may emerge from the enhancement required for a more efficient analysis of retinal images. Here we reassess the relevance of low-level vision tasks in the explanation of the Contrast Sensitivity Functions (CSFs) in light of (1) the current trend of using artificial neural networks for studying vision, and (2) the current knowledge of retinal image representations.}

\blue{As a first contribution, we show that a very popular type of convolutional neural networks (CNNs), called autoencoders, may develop human-like CSFs in the spatio-temporal and chromatic dimensions when trained to perform some basic low-level vision tasks (like retinal noise and optical blur removal), but not others (like chromatic adaptation} \green{or pure reconstruction after simple bottlenecks). As an illustrative example, the best CNN (in the considered set of simple architectures for enhancement of the retinal signal) reproduces the CSFs with an RMSE error of 11\% of the maximum sensitivity.}

\blue{As a second contribution, we provide experimental evidence of the fact that, for some functional goals (at low abstraction level), deeper CNNs that are better in reaching the quantitative goal are actually worse in replicating human-like phenomena (such as the CSFs).
This low-level result \green{(for the explored networks)} is not necessarily in contradiction with other works that report advantages of deeper nets in modeling higher-level vision goals.
However, in line with a growing body of literature, our results suggests another word of caution about CNNs in vision science since the use of simplified units or unrealistic architectures in goal optimization may be a limitation for the modeling and understanding of human vision.}

\vspace{-0.3cm}

}

\author{Li}{Qiang}
 {Image Processing Lab, Parc Cientific}
 {Universitat de València, Spain}
 {http://}{qiang.li@uv.es}
\author{Gomez-Villa}{Alex}
 {Computer Vision Center}
 {Universitat Autònoma de Barcelona, Spain}
 {http://}{alexander.gomez@upf.edu}
\author{Bertalmío}{Marcelo}
 {Instituto de Óptica}
 {Spanish National Research Council (CSIC), Spain}
 {http://}{marcelo.bertalmio@csic.es}
\author{Malo}{Jes\'us}
 {Image Processing Lab, Parc Cientific}
 {Universitat de València, Spain}
 {http://isp.uv.es}
 {jesus.malo@uv.es}

\keywords{Spatio-temporal and chromatic CSFs, Autoencoders,
\blue{eye MTF},
\blue{Noisy Cones}, Deblurring and Denoising, \blue{Chromatic Adaptation}, Bottlenecks, Natural Images, Statistical goals, Architectures. \textbf{Accepted in the Journal of Vision (special issue on \emph{Deep Neural Networks and Biological Vision})}}

\maketitle

\section{1. Introduction}

\textbf{The human Contrast Sensitivity Function (CSF)} characterizes the psychophysical response to visual gratings of different frequency~\cite{Campbell68}.
Filter characterizations in the Fourier domain are complete only for linear, shift-invariant systems.
Human vision certainly is more complicated than that, however, this simple measure of the bandwidth of the system is still of paramount significance in biological vision: the CSF filter is an image-computable model that roughly describes the kind of visual information which is available for humans~\cite{Watson16}.
Moreover, while it is defined for threshold conditions, there are many examples that illustrate
the relevance of the CSF in more general situations~\cite{Watson86,Watson02,Watson05},
so it has shaped image engineering over decades~\cite{Mannos74,Hunt75,Wallace92,Marcellin01}.
This theoretical and practical relevance motivated the measurement of CSFs, not only for spatial gratings~\cite{Campbell68}, but also for moving gratings~\cite{Kelly79}, chromatic gratings~\cite{Mullen85},  spatio-temporal-chromatic gratings~\cite{Mara11}, at different luminance levels~\cite{Wuerger20}, and for alternative basis of the image space~\cite{Malo97}.

\textbf{Principled explanations of the human CSFs.} \blue{Of course, the psychophysical CSFs have physiological roots in the spatio-temporal bandwidths of the center-surround cells tuned to achromatic and chromatic stimuli~\cite{Enroth66,Devalois71,Ingling83,Uriegas94,DeAngelis97,Reid92,Reid02}. However, the physiological basis of psychophysical phenomena does not explain the functional role (or \emph{goal}) of the underlying computation~\cite{Marr76,Marr82}.
The discussion about the \emph{goal} of certain mechanism relies on deriving the biological behavior from a computational principle. In the specific case of the CSFs, the classical work of Atick et al.~\cite{Atick92,Atick92a,Atick92b} derived the spatio-chromatic CSFs from the maximization of the information transferred from the input to the response of the system, which, under certain conditions is equivalent to optimal deblurring and denoising of the retinal signals.
These classical explanations were based on clever observations about the 2nd order properties of natural images, but relied on linear filtering models. As a result, the consideration of more flexible (non-linear) models could lead to a better fulfilment of the computational goal and, eventually to better explanations of the CSFs. A step forward in a more general (non-linear) derivation of these phenomena
from low-level principles was given by~\cite{Karklin11}, where they obtained sensors with center-surround receptive fields optimizing the information transferred by a linear+nonlinear layer of neurons with noisy inputs. However, this work did not considered the chromatic nor the temporal dimensions of the problem, and no explicit comparison with the psychophysical CSFs was done.
Similarly, \cite{LindseyICLR19} also reproduced center-surround sensors close to the retina when training anatomically constrained artificial neural nets (in this case training for a higher-level task such as object recognition).
Again, these center-surround cells eventually would induce CSFs, but this was not analyzed in that paper.}

\textbf{Emergence of CSFs in artificial neural networks.}
\blue{Automatic differentiation~\cite{Baydin18} has simplified the search of computational principles in vision science because it allows the optimization of complex models according to different goals without the burden of obtaining the analytical derivatives of the goals wrt the model parameters.
Automatic differentiation is at the core of the current explosion of \emph{deep-learning}~\cite{Goodfellow16}.
Full analytical description of the derivatives of realistic nonlinearities in visual neuroscience 
is certainly possible~\cite{Martinez18}, but the widespread availability of deep-learning tools
for simplified neurons makes the exploration of these artificial architectures much easier.
Conventional Convolutional Neural Networks (CNNs) are too simplistic from the neuroscience perspective\footnote{For example neurons in conventional CNNs have fixed nonlinearities, as opposed to the known adaptive nature of real neurons~\cite{Carandini12,Wilson73}}, but the freedom to combine multiple of such simplified layers in any possible way may compensate this shortcoming. In the end, one has a flexible system that can be optimized with automatic differentiation to fulfil whatever computational goal under consideration. As a result, deep-learning models are becoming standard in visual neuroscience~\cite{Kriegeskorte15,Yamins16,Cadena19}.}

\blue{According to the above, the study of the CSF of artificial neural networks is interesting for two reasons: (1) CNNs are flexible and easily optimizable tools which may allow us to investigate principled explanations of the human CSFs with more generality than the classical methods considered above, and (2) given the widespread use of CNNs in computer vision and its recent use in visual neuroscience, the eventual emergence of human-like sensitivities in these artificial systems has intrinsic interest.}

\blue{Very recently, two groups have reported complementary results on the emergence of CSFs
in deep networks:
first, in order to explain the human-like nature of some of the brightness and color illusions in CNNs trained for \textbf{\emph{low-level}} visual tasks found in~\cite{Gomez18}, a novel eigenanalysis of the networks was proposed~\cite{Gomez20b}. This analysis revealed the emergence of human-like chromatic channels, and achromatic and chromatic CSFs in these channels. Then, \cite{Arash21} have found that networks trained for \textbf{\emph{high-level}} visual tasks, such as classification, also may develop an achromatic CSF, in this case not explicitly imposing low-level constraints.}

\vspace{0.1cm}
\textbf{Contributions and scope of this work:}
\vspace{-0.1cm}
\begin{itemize}
\item \blue{\textbf{First}, following~\cite{Atick92,Atick92a,Atick92b,Karklin11} here we reconsider principled explanations of the CSFs from \textbf{\emph{low-level}} visual tasks in light of new available methods: (i)~the current tools from deep-learning, and (ii)~the current knowledge of retinal image representations. We check the emergence of spatio-temporal-chromatic CSFs in a wider range of low-level (goal/architecture) situations with more realistic inputs.}

\blue{Regarding the retinal input, we use recent models of the human Modulation Transfer Function (MTF)~\cite{Watson13}, and recent calibrated estimations of the noise in the cones~\cite{JoVnoise} obtained via the retina models implemented in ISETbio~\cite{Wandell19,Wandell20}.
In this way, here we generate realistic spatio-temporal noisy inputs to the visual pathway in a plausible representation: the cones of~\cite{Stockman00} tuned to \emph{Long}, \emph{Medium} and \emph{Short} wavelengths (LMS cones).}

\blue{Regarding the deep-learning tools, we use spatio-temporal extensions of the \emph{convolutional autoencoders} used in our analysis of color illusions in CNNs~\cite{Gomez18,Gomez20b}.
We elaborate on the proper determination of the CSF for convolutional autoencoders:
instead of the linear characterization of the autoencoder used in~\cite{Gomez20b}, which hides its nonlinear nature into a single matrix, here we stimulate the networks with gratings of different contrast. In this way the changes of the attenuation functions describe the nonlinearities of the system.}

\blue{Regarding the architectures, in this work we focus on autoencoders that reconstruct the signal in the input domain as opposed to the consideration of more general architectures that encode the images into more abstract representations to achieve higher-level tasks such as classification.
This limitation in scope is reasonable if one wants to model early vision stages like the Lateral Geniculate Nucleous (LGN) which do not imply change of domain and may function according to error minimization and signal enhancement principles~\cite{MartinezNeuron14}. If the CSFs are related to the response of LGN neurons as is usually assumed, autoencoders seem a reasonable computational framework to use.}

Regarding the tasks, in this \emph{low-level} context with autoencoder tools, we consider different visual tasks which may be implemented as early as in the retina-LGN path:
(a)~the enhancement of the retinal signal (related to information maximization) when the input is subject to different degrees of degradation due to different pupil diameters or different plausible levels of retinal noise,
(b)~the compensation of changes in the spectral illumination of the scene in a reasonable range of color temperature, and
\green{(c)~the reconstruction of the signal when some information may be lost in eventual bottlenecks.}


\item  \textbf{Second}, here we provide experimental evidence of \emph{"deeper CNNs are not necessarily better" (in representing this abstraction level)}. The bigger generality of flexible CNN models over fixed linear models is obvious, but one may ask: \emph{do more flexible architectures necessarily lead to more human CSFs?}, or \emph{does better accuracy in the goal imply more human CSFs and masking behavior?}. Consistently with previous results in low-level tasks~\cite{Gomez20b,Flachot20}
our CSF results presented below also seem to favour shallow networks \green{(in the explored range of architectures)}.

\blue{Our findings at this low-level of abstraction complement other results where deeper architectures actually imply closer resemblance to human behavior~\cite{Yamins14,LindseyICLR19,Cichy16,Cadena19}.
But this is not contradictory since they refer to different abstraction levels (high-level object recognition versus our low-level -color constancy and error minimization- goals).}

\end{itemize}

The structure of the paper is as follows.
Section~2 extends the theory proposed in~\cite{Gomez20b} to obtain the CSFs of autoencoders with an analysis of the energy (or standard deviation) of the input and the output gratings.
Section~3 describes the considered low-level visual tasks (compensation of bio-distortions, chromatic adaptation, \green{and signal reconstruction after bottlenecks}) and the setting of the numerical experiments.
Section~4 shows the main empirical findings of the work: the emergence of human-like CSFs in the spatio-temporal and chromatic dimensions in shallow CNN autoencoders trained to minimize the distortion introduced by the optics of the eye and the noise in the cones.
Finally, Section~5 discusses the implications of the empirical results: on the one hand, statements about the goal or organization principles are difficult to
separate from the implementation because the final behavior very much depends on the algorithmic level (or selected architecture).
On the other hand, special care has to be taken in using deep models in low-level vision science: their ability for function approximation may make them excel in the performance of a sensible score, but without the appropriate architecture constraints, this does not guarantee the similarity with humans.
Appendix~A provides details of the implementation of the models.
Appendix~B describes the image/video datasets to train the models and the sinusoids used to probe the networks.
Appendix~C illustrates the proper training and convergence of all the considered CNNs in all experiments. It shows the learning curves and explicit examples of the responses (reconstructed signals in test) for all the considered goal/architecture scenarios.

\vspace{0.4cm}
\section{2. Methods: Estimating Contrast Sensitivity in Autoencoders}
\label{Method}

Here we consider different linear characterizations of the autoencoders including the eigenanalysis proposed in~\cite{Gomez20b}.
That theory is extended with the explicit consideration of the image acquisition process in the human eye, which leads us to propose a procedure to estimate the autoencoder CSF that is more connected to the definition of the CSF in human observers.

\vspace{0.0cm}

\subsection{Autoencoders}

Autoencoders are artificial networks that transform the signal into an inner representation through an \emph{encoder}, and a \emph{decoder} transforms this inner representation back in the input domain.
\begin{equation}
\vect{x} \xrightarrow{\,\,\,\,N_\theta(\vect{x})\,\,\,\,} \vect{y}
  \label{autoencoder}
\end{equation}

In Eq.~\ref{autoencoder} we do not made explicit the encoding and decoding operations, i.e. $\vect{x}$ and $\vect{y}$ are in the image space.
In this work we will not make any assumption on the nature of the inner representation of the autoencoder.
This is because the basic goal function in autoencoders (reconstruction error) is defined in the image domain, shared by input and response.
Moreover, with the appropriate stimuli, the CSF characterization can be defined in this image domain.

Following~\cite{Gomez18,Gomez20b} we focus on Convolutional Autoencoders.
Below we will discuss and explore different  computational goals, but, for now, lets consider that the parameters $\theta$ are trained to compensate the blur and noise introduced in the signal by image acquisition process. In this context, given a clean image, $\vect{x}_c$, the input to the neural system would be a distorted version:
$\vect{x} = H \cdot \vect{x}_c + \vect{n}_r$, where $H$ is a blurring operator related to the optics of the eye and $\vect{n}_r$ is the noise associated to the response of the LMS photodetectors at the retina. Both $H$ and $\vect{n}_r$ are unknown to the neural system.
The goal of the network at this early stage is inferring $\vect{x}_c$ from $\vect{x}$.
Accurate models of LGN cells show that this may be one of the goals of the biological processing after retinal detection~\cite{MartinezNeuron14}.

In the supervised learning setting of ANNs, the parameters $\theta$ of the network are selected so that the average reconstruction error
$\varepsilon = |\vect{x}_c - N_\theta(\vect{x})|_2$ is minimized over a set of training images~\cite{Goodfellow16}.
\green{In our case, we will refer to the average reconstruction error as $\varepsilon_{\textrm{LMS}}$ since the input signal is expressed in the LMS cone space~\cite{Stockman00}}.
Of course, supervised learning and parameter updates using backpropagation may not be biologically plausible~\cite{AgainstHinton}.
However, our initial aim here is looking for statistical explanations of human frequency sensitivity and hence ANNs can be seen as convenient tools to optimize the selected goal.
With this focus on the \emph{goal}, the specific learning algorithm is not as important as ensuring that the final network actually fulfills the goal. We will see that the situation may not be that simple because networks optimizing the same goal with equivalent performance may display \emph{human} or \emph{non-human} CSFs depending on their architecture.

\subsection{Filter definition of the CSF and Linearized Autoencoders}

\afterpage{
\begin{figure}[b!]
\begin{center}
\begin{centering}
\hspace{-0cm}\includegraphics[width=0.75\linewidth]{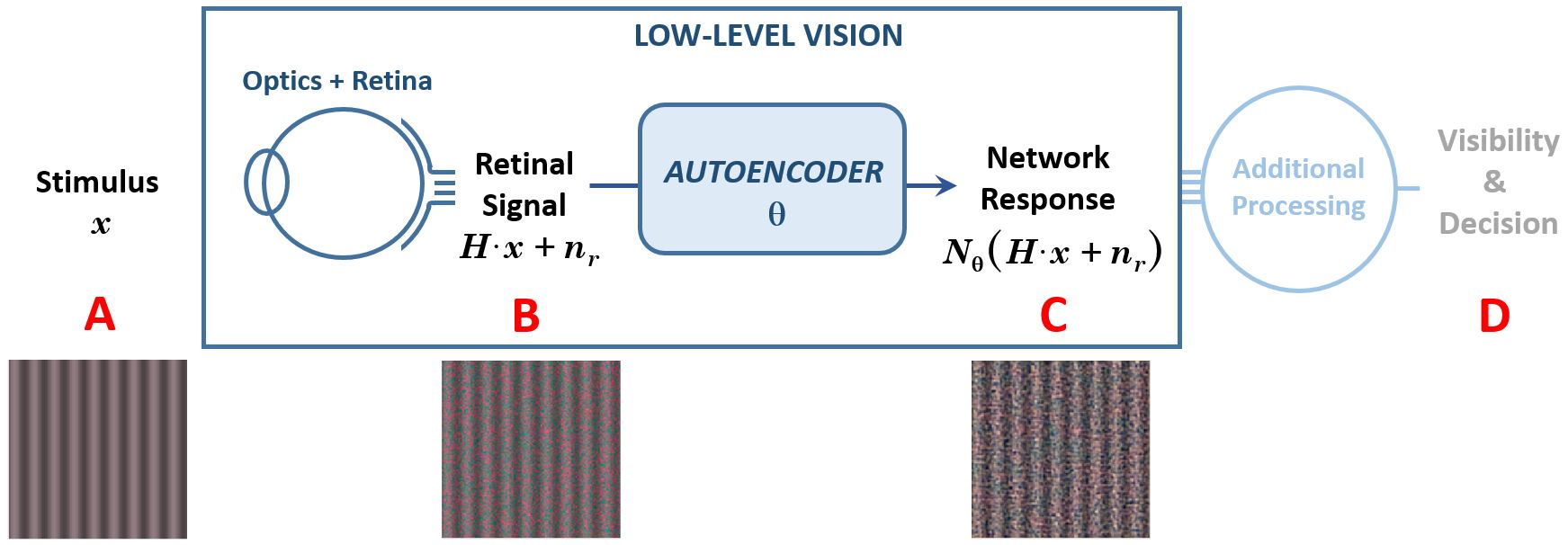}
\end{centering}
\vspace{-0.2cm}
   \caption{\textbf{Definition of the CSF as a frequency-dependent attenuation factor in a system to develop low-level vision tasks.} \blue{The diagram illustrates transforms in the visual signal from the input stimulus (A), the degraded signal due to optical blur and retinal noise (B), the process of the early neural path where the output is still in a spatial LMS representation (C) -modeled here by autoencoders-, and additional mechanisms that compute a decision on the visibility (D).
   In the conventional view of the CSF as a filter, the process from A to C is assumed to be linear and (in humans) the visibility of gratings is assumed to be based on the amplitude of the response at the point C. In human psychophysics (with no access to C) the observer makes visibility decisions, and the attenuation factors are determined from the thresholds. When dealing with artificial systems we do have access to the response in C so we do not need to model the decision mechanism and we can simply estimate the CSF from the ratio in Eq.~\ref{csf1}.}}
   \label{FilterDefinition}
\end{center}
\end{figure}
}

The CSF describes the linear response of human viewers for low-contrast sinusoids~\cite{Campbell68,Mullen85,Kelly79}.
In that linear setting, the CSF describes an input-output mapping where an input sinusoid of frequency $f$, the basis function $\vect{b}^f$, leads to an output, $\vect{y}^f$, with attenuated contrast (or attenuated standard deviation, $\sigma$).
The output standard deviation is given by, $\sigma(\vect{y}^f) = \textrm{CSF}(f) \, \sigma(\vect{b}^f)$. In the case of humans the attenuation factor, $\textrm{CSF}(f)$, has to be obtained from contrast thresholds because there is no access to the output. However, for autoencoders the computation of the output is straightforward.
\blue{If the degradation of the acquisition is taken into account,  the sinusoids, $\vect{b}^f$, used to simulate the measurement of the CSF have to undergo the degradation as well, and we should consider an \emph{eye+network} system, $S$:}
\vspace{-0.3cm}
\blue{\begin{equation}
  \xymatrixcolsep{6pc}
  \xymatrix{ \vect{b}^f \ar@/^2pc/[rr]^{\scalebox{0.90}{$\,\,\,\,\,S_\theta(\vect{b}^f)\,\,\,\,$}} \ar[r]^{\,\,\,\, H\cdot\vect{b}^f+\vect{n}_r \,\,} & \vect{b}^f_{\star} \ar[r]^{\,\,N_\theta(\vect{b}^f_{\star})\,\,\,\, } & \vect{y}^f}
  \label{eye_network}
\end{equation}
}

\vspace{-0.5cm}
Therefore, one could check the attenuation factor by comparing the standard deviation of output and input:
\begin{equation}
       \textrm{CSF}(f) = \frac{\sigma(\,\,S_\theta(\vect{b}^f)\,\,)}{\sigma( \vect{b}^f )} = \blue{\frac{\sigma(\,\,N_\theta(H \cdot \vect{b}^f + \vect{n}_r)\,\,)}{\sigma( \vect{b}^f )}}
       \label{csf1}
\end{equation}
\blue{Note that the CSF ratio in Eq.~\ref{csf1} (which uses degraded sinusoids to probe the network) is different from checking the Fourier response of the network, where one would use \emph{clean} sinusoids at the input:}
\begin{equation}
       \blue{\mathcal{N}^F(f) = \frac{\sigma(\,\,N_\theta(\vect{b}^f)\,\,)}{\sigma( \vect{b}^f )}}
       \label{csf_net}
\end{equation}
\blue{The relation of Eq.~\ref{csf1} with the regular determination of the CSF in humans is illustrated in Fig.~\ref{FilterDefinition}.}
Of course, the ratio in Eq.~\ref{csf1} should be computed for low-contrast sinusoids to keep parallelism with human CSF and keep the (eventually) nonlinear autoencoder in the low-energy range.
For chromatic sinusoids the deviations have to be computed separately over the achromatic, red-green, and blue-yellow color channels~\cite{Mullen85}.
In this work we use a classical opponent color space~\cite{Jameson59} to generate achromatic and purely chromatic gratings and to decompose the corresponding responses.

Of course, plain attenuation for sinusoids in Eq.~\ref{csf1} may not provide a full description of the action of nonlinear systems.
In principle it is not obvious why we should perform the analysis in a specific basis. Therefore one should check to what extent waves are indeed eigenfunctions of the system.

A way to test this point is linearizing the response of the autoencoders in the low-contrast regime and check that it is shift invariant.
Using a Taylor expansion, the response for low-contrast images can be approximated by the Jacobian around the origin (the zero-contrast image, $\vect{0}$, which is just a flat gray patch):
\vspace{-0.5cm}
\begin{eqnarray}
      \vect{y} & = & S_\theta(\vect{x}) \nonumber \\
      \vect{y}_{\textrm{low}} & = & S_\theta(\vect{0} + \vect{x}_{\textrm{low}}) \approx S_\theta(\vect{0}) + \nabla_{\!\vect{x}} S_\theta(\vect{0}) \cdot \vect{x}_{\textrm{low}} \\
      \vect{y}_{\textrm{low}} & \approx & \nabla_{\!\vect{x}} S_\theta(\vect{0}) \cdot \vect{x}_{\textrm{low}} \nonumber
\end{eqnarray}
\noindent where we assumed that the response for zero-contrast images is zero.
If the behavior of the system at this low-energy regime is shift invariant, the Jacobian matrix can be diagonalized as $\nabla_{\!\vect{x}} S_\theta(\vect{0}) = B\cdot \lambda \cdot B^{-1}$, with extended oscillatory basis functions in the columns of $B$ (and rows of $B^{-1}$).
Fourier basis and cosine basis are examples of extended (non-local) oscillatory functions that diagonalize shift invariant systems.
The reason for this result is equivalent to the emergence of cosine basis
when computing the principal components of stationary signals (shift invariant autocorrelation)~\cite{Clarke81}.
As a result, the slope of the response for low-contrast sinusoids (the CSF) will be related to the eigendecomposition of the Jacobian of the system at $\vect{0}$. Let's compute the response for a sinusoid in this Taylor/Fourier setting to see the relation.
A basis function $\vect{b}^f$ with specific frequency $f$ is orthogonal to all rows (sinusoids) in $B^{-1}$ except that of the same frequency, i.e. $B^{-1} \cdot \vect{b}^f = \vect{\delta}^{f'f}$. And this delta selects the corresponding column (of frequency $f$) among all the columns in the matrix B:
\vspace{-0.4cm}
\begin{eqnarray}
      \vect{y}^f & = & S_\theta(\vect{b}^f) \,\, \approx \,\,  \nabla_{\!\vect{x}} S_\theta(\vect{0}) \cdot \vect{b}^f \nonumber \\
                 & \approx & B\cdot \lambda \cdot B^{-1} \cdot \vect{b}^f \nonumber \\
                 & \approx & B\cdot \lambda \cdot \vect{\delta}^{f'f} \nonumber \\
                 & \approx & \lambda_f \, \vect{b}^f
                 \label{eq_eigen}
\end{eqnarray}
So the slope of the response for basis functions of frequency $f$ is $\lambda_f$ (the corresponding eigenvalue of the Jacobian of the autoencoder).
As a result, for systems with shift invariance in the low-contrast regime, the eigenvalues of the linear approximation of the system (eigenvalues of the Jacobian) are \emph{conceptually} similar to the CSF.
Direct comparison of the eigenvalue spectrum with the CSF may not be simple because the eigenfunctions may differ from Fourier sinusoids. Examples of this include isotropic systems (with a constant sensitivity for certain $|f|$ independent of orientation). In this case the eigenbasis may be not sinusoids but arbitrary linear combinations of sinusoids of the same frequency and different orientation.

Nevertheless, if the linearized version of the system (the Jacobian at $\vect{0}$) is shift invariant, which can be seen from a convolutional structure in the Jacobian matrix,
oscillatory waves are eigenfunctions of the system, and hence Eq.~\ref{csf1} may provide a good description of the behavior of the system.

\subsection{Alternative linear characterizations of the autoencoders}

\begin{figure*}[t]
\begin{center}
\includegraphics[width=0.8\linewidth]{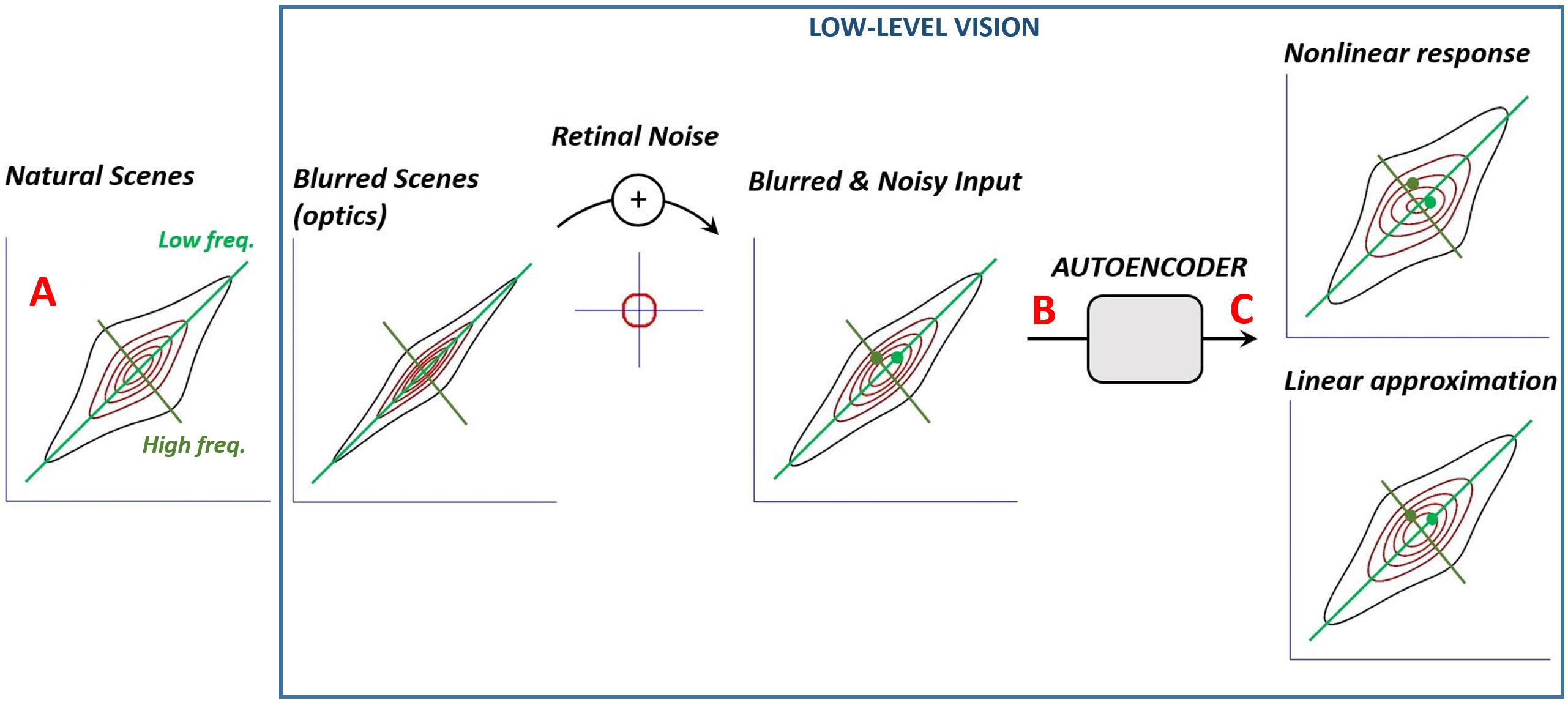}
   \vspace{-0.2cm}
   \caption{\textbf{Degradation of the signal PDF and linear and nonlinear strategies to compensate the degradation}.
   The axes of the plots represent the luminance in each of the photodetectors of two-pixel images as in \azul{(Simoncelli \& Olshausen, 2001)}.
   The left panel (A) represents the PDF of natural scenes: marginal heavy tailed distributions of oscillatory functions
   mixed to have strong correlation in the pixel domain. Optical blur (second panel) implies contraction of the high
   frequencies. Additive retinal noise implies the convolution by the PDF of the noise leading to the PDF in the third panel (B).
   The solutions to the restoration problem at point C may include (i) nonlinear transforms such as the one represented by the PDF at the right-top, and (ii) linear transforms as the one at the right-bottom.}
   \label{fig_degrad_restore}
   \vspace{-0.5cm}
\end{center}
\end{figure*}

A 2D cartoon of the impact of the degradation and restoration processes in the probability density (PDF) of the signal can illustrate alternative characterizations of the neural networks optimized to enhance the retinal signal (see Fig.~\ref{fig_degrad_restore}).
In this diagram, two-pixel natural scenes (the left panel) follow a PDF obtained from independent t-student sources mixed by a matrix that introduces strong correlation between the luminance of the pixels.
This kind of two-pixel representations is common to describe the statistics of natural images~\cite{Olshausen01},
and mixtures of sparse components is a widely accepted model for natural scenes~\cite{Hyvarinen09,JMLR14,Malo20}, and
appropriate enough for this illustration.
In this diagram the low-frequency direction corresponds to the main diagonal (where the two pixels have the same luminance)
and the high-frequency direction is orthogonal (for images where one of the pixels is brighter than the other).
The zero-contrast image is at the crossing point of the frequency axes.

Optical blur implies the attenuation of high-frequency versus low-frequency components and hence the contraction of the dataset
as shown in the second panel. Assuming linear and noisy photoreceptors, the PDF of the retinal response results from the convolution
of the PDF of the blurred images with the PDF of the noise (the function with circular support).
The result (third panel) is the input to the autoencoder, whose goal is recovering the distribution at the first panel.
Linear solutions are limited to global scaling of the domain (for instance by inverting the contraction introduced by the blur),
while nonlinear solutions may twist the domain in arbitrary ways.

In this setting, the computation of the CSF according to Eq.~\ref{csf1} means putting low-contrast sinusoids (e.g. the samples
highlighted in green in the third panel) through the system, and checking the amplitude of the output (green dots at
the panels at the right) over the directions of the input.
This nonlinear example illustrates the fact that the behavior can be contrast dependent (see the different twist in the concentric contours).
This graphical view illustrates the difference between three possible linear characterizations, with $\vect{y} = M\cdot \vect{x}$:
\vspace{0.0cm}
\begin{itemize}
     \item \textbf{The optimal linear solution:} the matrix $M$ that better relates the input $\vect{x}$ with the desired output $\vect{x}_c$.
           This is the $M$ that minimizes the expected value
           $E\{|\vect{x}_c - M \cdot \vect{x}|_2\}$.
           Assuming a representative set of $N$ clean/distorted pairs stacked in the matrices $\vect{X_c} = [\vect{x}_c^{(1)} \vect{x}_c^{(2)} \ldots \vect{x}_c^{(N)}]$ and $\vect{X} = [\vect{x}^{(1)} \vect{x}^{(2)} \ldots \vect{x}^{(N)}]$, the optimal solution in Euclidean terms is given by the pseudoinverse:
           \begin{equation}
                  M = \vect{X_c} \cdot \vect{X}^\dag
                  \label{linear_solution}
           \end{equation}
     \item \textbf{Globally linearized network:} the matrix $M$ that better describes the nonlinear behavior of the
           network \emph{over the whole set of natural images}.
           This is the $M$ that minimizes $E\{|S(\vect{x},\theta) - M \cdot \vect{x}|_2\}$.
           Assuming a representative set of $N$ input/output pairs stacked in the matrices $\vect{X} = [\vect{x}^{(1)} \vect{x}^{(2)} \ldots \vect{x}^{(N)}]$, and $\vect{Y} = [\vect{y}^{(1)} \vect{y}^{(2)} \ldots \vect{y}^{(N)}]$, the solution is given by the pseudoinverse:
           \begin{equation}
                  M = \vect{Y} \cdot \vect{X}^\dag
                  \label{global_linear_approx}
           \end{equation}
     \item \textbf{Locally linearized network at $\vect{0}$:} the matrix $M$ that better describes the nonlinear behavior of the
           network \emph{for low-contrast images}. This is the $M$ that minimizes $E\{|\nabla_{\!\vect{x}}S(\vect{0},\theta)\cdot\vect{x}_{low} - M \cdot \vect{x}_{low}|_2\}$. Of course, this could be empirically approximated by $M = \vect{Y}_{low} \cdot \vect{X}_{low}^\dag$, but in this case
           the obvious exact solution is:
           \begin{equation}
                  M = \nabla_{\!\vect{x}}S_\theta(\vect{0})
                  \label{local_linear_approx}
           \end{equation}
\end{itemize}

While the optimal linear solution (or the optimal linear network) is a convenient reference to describe the problem, the other two options are different characterizations of the autoencoder. Eq.~\ref{global_linear_approx} summarizes the behavior of the network in a single matrix,
and Eq.~\ref{local_linear_approx} is a description only valid around $\vect{0}$, and hence more closely connected to the low-contrast regime of the CSF. The eigenanalysis cited for $\nabla_{\!\vect{x}}S$ in Eq.~\ref{eq_eigen} can be applied for the three matrix characterizations, but it is important to note the differences between them.

The Jacobian of cascades of linear+nonlinear layers (as in autoencoders based on Convolutional Neural Networks) can be obtained
either
analytically\footnote{For optical blur where the linear operator $H$ can be obtained from the MTF~\cite{Watson13}, and the retinal noise is Poisson,
$\vect{n}_r = F \cdot \mathbb{D}_{\left( |H \cdot \vect{x}|^{\frac{1}{2}} \right)} \cdot \vect{n} $, where $\mathbb{D}_{\vect{v}}$ is a diagonal matrix with vector $\vect{v}$ in the diagonal, $F$ is the Fano factor, and $\vect{n}$ is drawn from a unit-variance Gaussian~\cite{JoVnoise};
the Jacobian in Eq.~\ref{local_linear_approx}, is  $\nabla_{\!\vect{x}}S_\theta(\vect{0}) = \nabla_{\!\vect{x}}N_\theta(\vect{0}) \cdot \left( I - \frac{F}{2} \cdot \mathbb{D}_{ \left( \vect{n}\odot|H\cdot \vect{0}|^{ \frac{1}{2}} \right) } \right) \cdot H$, where the Jacobian of the network, $\nabla_{\!\vect{x}}N_\theta(\vect{0})$, can be obtained analytically~\cite{Martinez18}.}, or it can be obtained via automatic differentiation or alternative methods
based on system identification~\cite{Laparra17}.
However, the above procedures are tedious, so in~\cite{Gomez20b} we took the more straightforward approach represented by Eq.~\ref{global_linear_approx}.

The different linear characterizations considered in this section and the diagram in Fig.~\ref{fig_degrad_restore} illustrate that
the behavior of a nonlinear autoencoder for high contrasts may be substantially different from the threshold behavior.
Therefore, the attenuation of sinusoids by the linearized system (by the matrix Eq.~\ref{global_linear_approx})
will be compared with the result of Eq.~\ref{csf1}.

\vspace{0.0cm}
\subsection{Limitation of the proposed CSF definition in autoencoders}

\blue{In order to maximize the equivalence to human CSFs, the proposed procedure (the ratio in Eq.~\ref{csf1} which compares the signals at points C and A in Fig.~\ref{FilterDefinition}) implies the consideration of the retinal degradation process. This consideration of the retinal noise will be shown to improve similarity with human CSFs in the Experiment~3 below, but it comes at a cost.
Note that even if the role of the autoencoder is compensating the retinal noise, complete removal is not possible. Therefore, there is some residual distortion in the response after the autoencoder. As a result, the standard deviation in the numerator of the proposed Eq.~\ref{csf1} not only measures the contrast of the output grating, but also measures the energy of the residual noise. In this way, when the contrast of the sinusoids $\vect{b}^f$ is very small, as expected in threshold conditions, the standard deviation maybe measuring more the residual noise than the contrast of the output. The limitation of Eq.~\ref{csf1} is that it has to be applied to sinusoids of relatively high contrasts so that the energy of the response coming from the sinusoid is bigger than the energy of the response coming from the noise.}

\blue{One can overcome this limitation in two ways: (1) by computing the response many times for different noise evaluations and cancelling the residual noise by averaging over the realizations, and (2) by using relatively high contrast sinusoids so that the effect of the residual noise is negligible.}

\blue{In this work (for computational convenience) we used the second approach: we probed the models with sinusoids with contrasts in the range [0.07, 0.6].
The lower limit is certainly higher than the minimum absolute threshold of the Standard Spatial Observer (which is about 0.005)~\cite{Watson02,Watson05}. Nevertheless, we choose this range for two reasons: first, 0.07 is the average of the threshold achromatic contrasts in the Standard Spatial Observer, and second, we empirically checked that the effect of the noise was negligible above this value.}

\vspace{0.4cm}
\section{3. Experiments}

The introduction raised questions on the role of low-level vision goals to explain the CSFs, the emergence of the CSFs in autoencoders working to solve these goals, and the eventual advantage of progressively more flexible models in explaining the CSFs.
In order to address these issues in the more general spatio-temporal-chromatic case,
\green{(1)~we perform two extensive experiments (one with images, and one with video) to compensate biologically sensible degradation of the retinal signal (compensation of bio-distortion), using a range of CNN architectures of different depth or flexibility,}
(2)~\green{we consider alternative low-level functional goals such as chromatic adaptation and the compensation of the effect of bottlenecks, }
\green{(3)~we consider different levels of bio-distortion, chromatic shifts in different directions, and bottlenecks with different restrictions, and
(4)~we consider the consistency of the results under changes in the statistics of the signal. In this section we describe the experimental setting of these simulations.}

\vspace{0.0cm}
\subsection{Functional goals}

\green{\textbf{Compensation of retinal bio-distortion (biological blur and noise):}} consists of overcoming the degradation introduced in the acquisition of the visual signal. Specifically,
the top panel in Fig.~\ref{fig_goals} shows how a natural scene is degraded at the output of the retina according to the variations of the eye MTF for different pupil diameters (from top to bottom, $d=2mm$, $d=4mm$ and $d=6mm$), and a sensible range of Poisson retinal noise levels (from left to right, Fano factors $F=0.25$, $F=0.5$ and $F=1$).
Variations of the MTF have been simulated with the expression in~\cite{Watson13},
and the noise in LMS sensors has been estimated in the discrete representation of the input digital image as in~\cite{JoVnoise}. In that work noise was obtained by stimulating the ISETBio retinal model~\cite{Wandell19,Wandell20} with flat stimuli of controlled size and tristimulus values over short and long exposure times. Cartesian resampling of the random cone mosaic of the retinal model and integration of the photocurrents over space/time reveals the effective Poisson nature of the noise (in the original LMS units) and allows the estimation of the effective Fano factor in the original discrete grid of the input image~\cite{JoVnoise}.
In that way we can easily generate calibrated noisy retinal images by adding this effective Poisson noise in the LMS representation of the digital image.
The illustrations in Fig.~\ref{fig_goals} come from the transformation of the LMS tristimulus images into the RGB digital counts for proper display.

\afterpage{
\begin{figure}[h!]
\begin{center}
\includegraphics[width=0.675\linewidth]{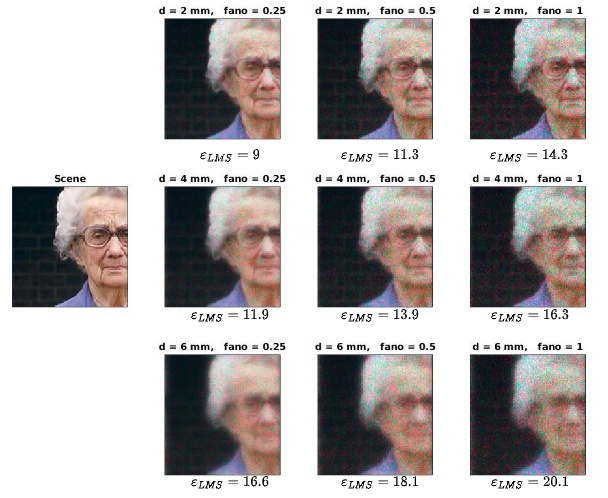}\\[0.75cm]
\includegraphics[width=0.9\linewidth]{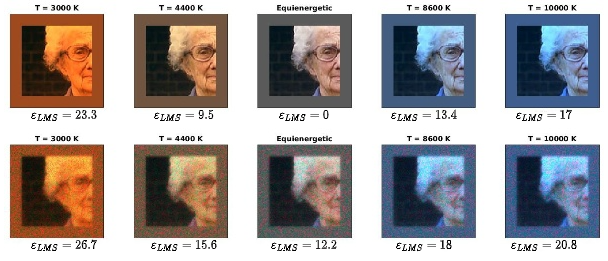}
   \vspace{-0.2cm}
   \caption{\textbf{Functional goals.}
   Possible low-level goals of the autoencoders are compensating the following distortions in the visual input. \emph{Top~panel:} different levels of retinal bio-distortion. \emph{Bottom~panel (first row)}: changes in the spectral illumination, \emph{Bottom panel (second row)}: changes in illumination + retinal bio-degradation. Original image from the ImageNet
   database~\azul{(Russakovsky et al. 15)}.
   An alternative low-level goal is the reconstruction of the signal in presence of bottlenecks (as in the architectures considered below, Fig.\ref{Archi} right).}
   \label{fig_goals}
\end{center}
\end{figure}
}


\green{\textbf{Chromatic adaptation:}} consists of the compensation of the deviations of the signal induced by the change of illuminant.
The bottom panel shows how the image of a natural scene changes under changes in the shape of the spectral radiance of the illuminant. Change of illuminant in a digital image was simulated in this way: each pixel of the image was associated with a reflectance chosen from a large database of natural reflectances so that under an equienergetic illuminant led to the tristimulus values of the pixel. Then, a black-body radiator, which simulates natural ambient light along the day, was used to generate spectra of the same energy but different shape. From there, we could get versions of the scene under arbitrary color temperatures. This process is straightforward using the functionalities and databases of Colorlab~\cite{Colorlab}. Of course, this process is just an approximation because it disregards the (unknown) geometry of the scene and assumes a flat´Lambertian world. Nevertheless, as illustrated in the examples of Fig.~\ref{fig_goals}, it does a good qualitative job to generate controlled samples to check chromatic adaptation in large image databases.

\green{\textbf{Compensation of chromatic shifts + bio-distortions.}} \blue{The reason to consider this combination is that pure chromatic shifts with no additional distortion is not a realistic input for the visual pathway:
the image acquisition front-end \emph{does exist} and hence what we called bio-distortion has to be taken into account.
Such combination of distortions is illustrated by the second row of the bottom panel in Fig.~\ref{fig_goals}.} Note that in the examples involving chromatic deviations (bottom panel) we introduced a flat-reflectance frame to help the models to cope with the chromatic adaptation~\footnote{We prepared the samples that way before actually knowing how well the networks are able to cope with this distortion.}.

\begin{figure}[t!]
\begin{center}
\includegraphics[width=1.0\linewidth]{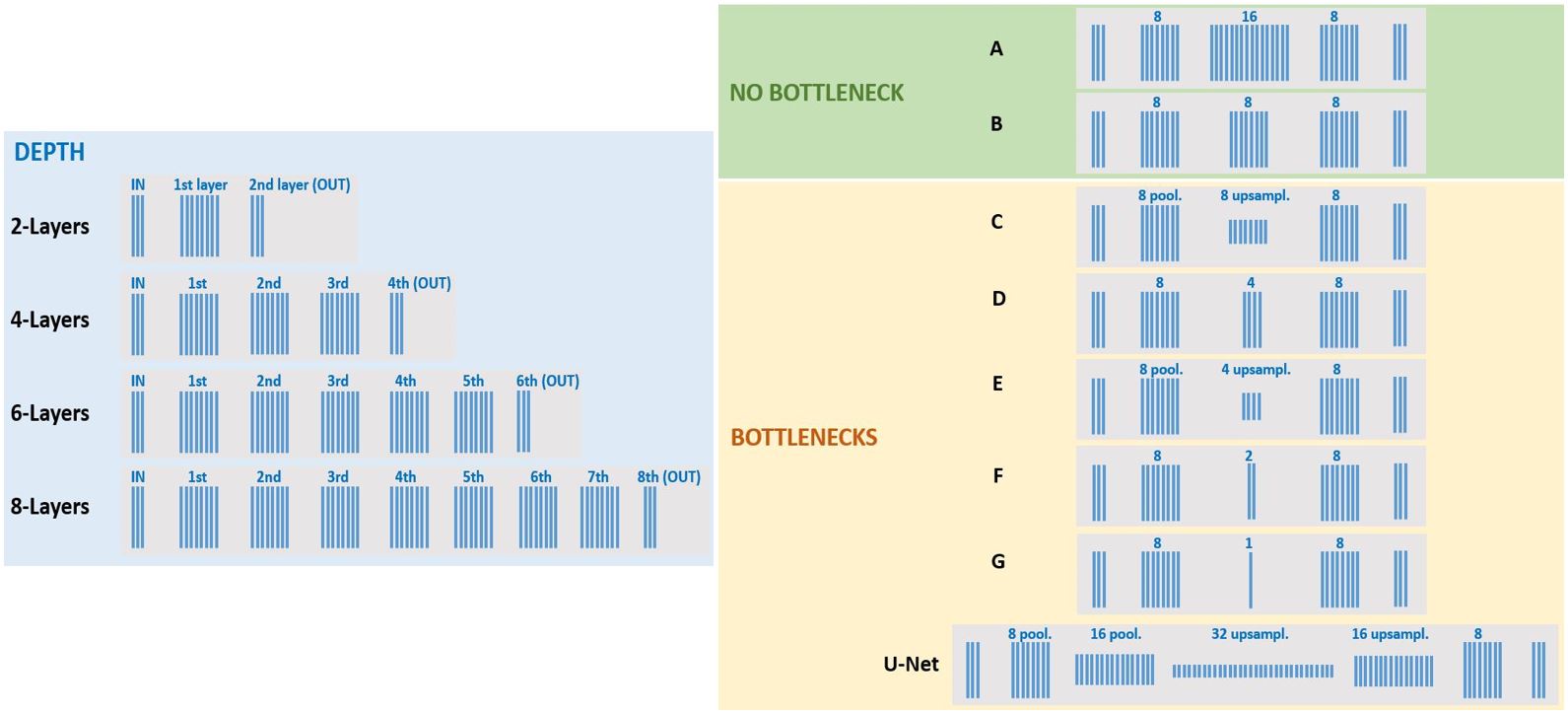}\\
   \vspace{-0.2cm}
   \caption{\textbf{Architectures.}
   \textbf{(Left)} Range of architectures of different \emph{depth} following the basic structure of the nets studied
   in~\azul{(Gomez-Villa et al. 20)}:
   three channels at the input and at the last (output) layer. In this work the input is in an LMS color representation. The rest of the layers have 8 features with no undersampling or bottleneck, as represented by the 8 blue lines of the same length.
   These architectures of increasing depth are used to study the compensation of the bio-distortion in images and videos.
   The best of these architectures in terms of CSFs (which turns out to be the 2-layers one) is used with images to explore chromatic adaptation.
   \textbf{(Right)} Range of architectures used to illustrate the effect of \emph{bottlenecks}. In this case, the inner layer in the 4-layer architecture at the left is systematically expanded (in A) or contracted (from C to G, either in the number of features or in the spatial resolution -see numbers and indications of pooling/upsampling and corresponding length of the layers-) to generate a range of bottlenecks. Finally, an illustrative U-Net with no residual connections is also considered in the experiments with bottlenecks.}
   \label{Archi}
\end{center}
\end{figure}

\textbf{Compensation of bottlenecks (pure reconstruction):} consists of
recovering the input after the signal has gone through a bottleneck.
Examples of bottlenecks include the restriction of the spatial resolution or the restriction of the number of
features (or channels) in the representation. Fig.~\ref{Archi} (right panel) shows an illustrative range of architectures:
from cases that expand the number of features (no bottleneck) to a variety of
cases that introduce local pooling, reduce the number of features, or try to compensate the effect of spatial undersampling by increasing the number of features. Bottlenecks may imply severe information loss if the representation is not optimized. Therefore, pure reconstruction of the signal in presence of bottlenecks is a sensible low-level goal to explore.

\textbf{Compensation of bottlenecks + bio-distortion.} As stated above, errors in the acquisition front-end (bio-distortion) do exist, so its consideration together with bottleneck compensation makes the goal more realistic.

\blue{All in all, we explored 9 levels of bio-degradation, chromatic adaptation in the blueish and the reddish directions (T = 8600~K and T = 4400~K respectively), and the combination of the central bio-distortion with the considered chromatic deviations.}
\green{We considered a pure reconstruction task with the 8 bottleneck configurations in Fig.~\ref{Archi}-right, and the compensation of these bottlenecks was also combined with the central bio-distortion case.}
The optical/retinal degradation in movies was applied in a frame-by-frame basis. No experiments involving chromatic adaptation or bottlenecks were done in movies, but only in natural and cartoon images.

\vspace{0.5cm}

The above computational goals are all measured in \emph{distortion} terms, \green{or how well the deviations $\varepsilon_{\textrm{LMS}}$ were compensated}. However, even within this low abstraction level, other computational goals could be considered together with the distortion, as for instance the \emph{information} or the \emph{energy} of the signal. In the experiments we restrict ourselves to the considered cases of distortion minimization and purely architectural bottlenecks.
The discussion suggests how the goals considered here could be related or combined with other kind of goals or more general (energy or information) bottlenecks.


\vspace{0.0cm}
\subsection{Architectures}

In this work we consider 2D CNNs that act on spatio-chromatic signals and 3D CNNs that act on spatio-temporal signals (color images and color videos).
The set of explored architectures is shown in Fig.~\ref{Archi}. These are variations of the basic toy networks studied in~\cite{Gomez18,Gomez20b}: autoencoders with convolutional layers made of 8 feature maps with kernels of spatial size $5\times5$ and sigmoids or Rectified Linear Units (ReLU) as activation functions. From that starting point, here we consider a range of nets of increasing \emph{depth} and flexibility: from the linear network in Eq.~\ref{linear_solution} (as a convenient base-line reference of 1 layer with no flexibility), and CNNs with 2 layers to 8 layers, both for the 2D and 3D cases.
Moreover, we also consider a range of architectures with different \emph{bottlenecks}, in this case, only 2D.

\blue{Of course, the range of possible architectures is virtually infinite and an exhaustive exploration of the architecture space is out of the scope of this work.
However, note that the considered set of architectures of \emph{progressive flexibility and constraints} is appropriate for the aim of this work for two reasons:
(1) these architectures do a good job in fulfilling the goal so they are good examples to reason about systems that work according to the considered function, and
(2) they display a range of flexibility and accuracy in the goal which is appropriate to illustrate the proposed questions (eventual emergence of the CSF and other nonlinearities, and qualitative effect in the CSF of increased flexibility and improvements in the goal accuracy).}

\blue{The first point (the considered \emph{toy} models do a reasonably good job in fulfilling the goal) is a technical issue that is demonstrated by the performance tables shown below and by the specific learning curves and reconstructions included in the Appendix A. However, to put this quantitative performance in context, it is interesting to note that the retinal bio-distortion is not an easy task to solve for general-purpose state-of-the-art image restoration CNNs.
In particular, following~\cite{Gomez20b}, on top of the described \emph{toy} networks, the computation of the CSFs of cutting-edge deeper models designed for restoration could be an illustrative limit to consider.
However, we found that the combination of representative examples of \emph{generic} CNNs for denoising~\cite{Zhang17,Soh21} and deblurring~\cite{Tao18}, which gave excellent results with arbitrary Gaussian noise and blur in~\cite{Gomez20b}, is not satisfactory
with biological distortion.
In particular, generic enhancement algorithms did not produce better results than the considered simple architectures (specifically trained for this bio-distortion).}
%
%
\blue{Of course, this does not mean that the toy models used here are better than the state-of-the-art, nor that state-of-the-art models are intrinsically unable to deal with this biological degradation.
One could certainly fine-tune these deep architectures for the bio-distortion and then get a better result than with the considered set of architectures, but that is not the goal of this work.
The relevant argument in favour of the considered (toy) architectures for our purposes here is this: the fact that generic blind restoration CNNs need to be retrained to get better results than the proposed models means that these simple models can be considered as good (enough) examples of systems actually fulfilling the goal.}

\blue{Regarding the second point (the considered set of architectures is good enough to illustrate interesting questions), consider that (i)~according to the results presented below (Section~4, Tables~1 and~4) the toy nonlinear models reduce up to 35\% and 48\% the error of the optimal linear solution in images and video respectively, and (ii)~the best nonlinear model reduces the error of the shallower nonlinear model by 21\% and 12\% in images and video respectively.}

\blue{In summary, the considered set of architectures
(progressively deeper CNNs and a range of bottlenecks)
does a reasonable job in optimizing the goals, and it is wide enough to illustrate changes in the achievement of the goals. As a result, the considered set of architectures is appropriate to address the questions raised in the introduction.}

See Appendix A for implementation details. Data and code are available at \azul{http://isp.uv.es/code/visioncolor/autoencoderCSF.html}

See Appendix B for details on the databases to generate the training stimuli and the stimuli used to probe networks.

\vspace{0.0cm}
\subsection{Assessing the quality of the CSF results}

\blue{The CSFs defined for the autoencoders may be subject to two arbitrary scale factors. On the one hand, the response of the network could be multiplied by an arbitrary global scale factor and hence, the numerator in Eq.~\ref{csf1} (and the CSF amplitude) would be multiplied by this scale factor as well. We will refer to this global scale factor in the amplitude as $\alpha_{\textrm{CSF}}$. On the other hand, the sampling assumptions (or assumptions on the extent of the signal, or the viewing distance) introduced in the description of the stimuli are arbitrary and they imply an arbitrary scaling in the frequency axis of our Fourier domains.
We will refer to this scale factor on frequency as $\alpha_f$.}

\blue{The factor on amplitude is not a major problem:
one network and a modified version with its outputs multiplied by $\alpha_{\textrm{CSF}}$ are equivalent and their quality should be rated the same. The factor on frequency does not reduce the validity of the results either as long as it is moderate.
Note that using the MTF expressions in~\cite{Watson13}, if the filter corresponding to a pupil of 3.5mm is modified by applying $\alpha_f=0.75$ or $\alpha_f=4.5$, the resulting MTF is similar to what would have been obtained with d=2mm or d=6mm respectively. Therefore, as changes in the MTF (the only element where the scaling in frequency matters)
are plausible if $\alpha_f \in [0.75, 4.5]$, one should also discount moderate variations of this factor when assessing the quality of the CSFs.
}

\green{The similarity between the model and the human CSFs will be measured by the Euclidean distance between the CSF vectors, averaging over the frequency, $f$, and the chromatic channels, $c$ (achromatic, red-green and yellow-blue), which will be referred to as:}
\begin{equation}
    \green{\textrm{RMSE} = \left( \sum_{f, \,c} \,\, (\,\, \textrm{CSF}_{c}^{\,\,\textrm{scaled}}(f) - \textrm{CSF}_{c}^{\,\,\textrm{human}}(f) \,\,)^2 \right)^\frac{1}{2}}
    \label{def_RMSE}
\end{equation}
where the \emph{scaled} attenuation factors of the model are related to the \emph{raw} attenuation factors of the model as:
\begin{equation}
    \blue{\textrm{CSF}^{\,\,\textrm{scaled}}(f) = \alpha_{\textrm{CSF}} \cdot \textrm{CSF}^{\,\,\textrm{raw}}( \alpha_f \cdot f)}
    \label{scaling}
\end{equation}
In the following we will report the scaled CSFs together with the scaling factors that minimize the distance with human CSFs.

\blue{It is important to mention that the relative scaling between the CSFs in the three chromatic channels is a characteristic feature of a network (or system) and it should not be modified. Therefore, the same factors in Eq.~\ref{scaling} are applied to the three CSFs. With these considerations, the CSFs reported below represent the closest approximation the models may give to the human CSFs, and hence the comparison between them is fair.}

\green{The magnitude of the RMSE errors has to be understood in reference to the maximum value of the human sensitivity.
As a convenient example to have in mind, RMSE = 22 corresponds to an average deviation of 10\% of the scale of the human spatio-temporal CSF at every frequency and chromatic channel. This is because the maximum sensitivity is about 200 for stationary gratings and about 220 for moving gratings~\cite{Watson02,Kelly79}.}



\vspace{0.0cm}
\subsection{List of experiments}

The empirical exploration of the considered architectures
consists of \emph{six} experiments. Experiments~1-5 deal with spatio-chromatic stimuli and 2D~networks, and Experiment~6 deals with spatio-temporal-chromatic stimuli and 3D~networks.
\green{As stated above, the computational goals are measured by the Euclidean distance between the reconstructed image and the original image, referred to as $\varepsilon_{\textrm{LMS}}$. The similarity with the human behavior is measured in terms of the Euclidean distance between the model CSFs and the human CSFs, i.e. the RMSE defined in Eq.~\ref{def_RMSE}.}

\begin{itemize}

    \item \textbf{Experiment 1: Spatio-chromatic CSFs from bio-distortion compensation by a range of architectures.}
    \blue{This experiment is focused on the central degradation shown in the first panel of Fig.~\ref{fig_goals} (d=4mm, F=0.5) and analyzes in detail the CSFs for nine architectures: the optimal linear network, and eight CNN architectures with 2-, 4-, 6-, and 8-layers with either sigmoid or ReLU activations, all optimized according to this distortion-compensation goal.
    Once the architectures are properly trained (using 20~$\cdot 10^3$ images of the ImageNet database cited in Appendix~B, 18~$\cdot 10^3$ for training and 2~$\cdot 10^3$ for validation), we get the numerical performance of the models in the independent test set of $10^3$ images. The sizes of the train/validation/test sets are the same in all experiments with images, Exp. 1 to 5. Throughout all the experiments, the performance is expressed as the average $\varepsilon_{\textrm{LMS}}$ of the reconstruction in LMS space over 20 batches of 50 randomly chosen images/batch. The standard deviation over these 20 computations is also reported.
    The learning curves (train/validation) and the reconstructions of one representative test image are given in Appendix~C.
    Then, the CSFs (attenuation factors) of the trained models are computed according to the method described in Section~2 for gratings of different contrasts. The eventual variation of the attenuation reveals the nonlinear nature of the contrast response for gratings.
    In Experiment 1 we also show the CSFs of the linear network and the linearized versions of the nonlinear networks introduced in Section~2.
    From the results of Experiment~1, one of the nonlinear models is chosen as having representative resemblance with human behavior in terms of the CSFs (2-layers with ReLU activation). Experiments~2, 3, and~4 further explore the behavior of this specific model in a number of conditions.}

    \item \textbf{Experiment 2: Consistency of the CSFs from bio-distortion compensation over a range of distortion levels.} \blue{This experiment is focused on the representative architecture selected after Experiment 1,
    and checks its CSFs when trained for the \emph{nine} different degradation levels considered in the first panel of Fig.~\ref{fig_goals}.}

    \item \textbf{Experiment 3: CSFs from chromatic adaptation and bio-distortion compensation.} \blue{This experiment checks the CSFs of the representative architecture selected after Experiment 1,
    when it is trained for (i)~the bio-degradation compensation alone, (ii)~the degradation compensation together with compensation of a bluish illuminant, (iii)~the degradation compensation together with compensation of a reddish illuminant, (iv)~pure compensation of a bluish illuminant, and (v)~pure compensation of a reddish illuminant.
    In the illustration of Fig.~\ref{fig_goals}, these correspond to the five distorted versions closer to the clean image under equienergetic illuminant.
    As stated above, the purely chromatic deviations are not realistic because they disregard the optics and retinal noise. However, they represent an illustrative reference.
    In the same vein, as a convenient reference, in this experiment we compute the CSF in two ways: (a)~the proposed (realistic) way, Eq.~\ref{csf1}, by putting the clean gratings through the retinal degradation before entering the network, and (b)~the idealized way, Eq.~\ref{csf_net}, in which we simply put the clean gratings through the considered network.
    This will stress the difference in the obtained CSFs when considering realistic spatial degradations or not.}

    \item \textbf{Experiment 4: Consistency of the human/non-human CSFs under change in signal statistics.} \blue{Here we reconsider the chromatic adaptation and the degradation-compensation goals of Experiment 3 now using stimuli of (apparently) quite different spatio-chromatic statistics: the images from the Pink Panther cartoons. All the other settings remain the same as in Experiment 3.}

    \item \textbf{\green{Experiment 5: CSFs from bottleneck-compensation and bio-distortion compensation.}} \green{This experiment shows the CSFs of the systems that emerge from imposing pure reconstruction of the signal in presence of bottlenecks in the network (the 8 examples in Fig.~\ref{Archi}, right). Pure reconstruction is compared with the compensation of bio-distortion in the same architectures. Given the similarity between activation options found in Experiment 1, here we just explore the ReLU case.}

    \item \textbf{Experiment 6: Spatio-temporal-chromatic CSFs from bio-distortion compensation by a range of architectures.} \blue{Here we check the fundamental findings of Experiment 1 for spatio-temporal-chromatic gratings on 3D~networks optimized for degradation-compensation.
    Given the similarity between activation options found in Experiment 1, here we just explore the sigmoid case.
    Therefore, we explored \emph{five} architectures: the linear one and 2,4,6, and 8 layers with sigmoid.
    In this spatio-temporal case we used 22~$\cdot 10^3$ video patches in the learning (20~$\cdot 10^3$ for training and 2~$\cdot 10^3$ for validation), and 3~$\cdot 10^3$ for test.
}
\end{itemize}

\vspace{0.4cm}
\section{4. Results}

\blue{Results in all the experiments have two parts:
(1) the \emph{perception} part, with the CSFs and the contrast responses of the networks, and (2) the \emph{technical} part, with evidences of the convergence of the models, numerical performance in reconstruction, and visual examples of the performance in reconstructing images.
The main text is focused on the perception part, while all the technical material is given in the Appendix~C.}

\subsection{Experiment 1: Spatio-chromatic CSFs from bio-distortion compensation}

\blue{Figure~\ref{Exp1_csfs} shows the achromatic and chromatic CSFs of the considered models (the linear solution and the eight CNNs) together with the human CSFs for convenient reference. The human data come from the achromatic standard spatial observer~\cite{Watson02,Watson05} and from the measurements in~\cite{Mullen85}).
The plots for the nonlinear models show the attenuation factors (CSFs) for gratings of different contrast (dark to light colors mean lower to higher contrasts).}

\blue{These plots include the RMSE measure of the difference of the artificial CSFs with the human CSFs.
The insets also show the optimal values of the arbitrary scaling factors $(\alpha_f, \alpha_{\textrm{CSF}})$ applied to the axes of the raw CSFs of the network to minimize the distance with the human CSFs. Since these optimal scaling factors values were found exhaustively in all cases, the comparison of the final CSFs and RMSE values is fair.}

\afterpage{
\begin{landscape}
\vspace{5cm}
\begin{figure}[h!]
\vspace{4cm}
   \centering
    \includegraphics[width=1.2\textwidth]{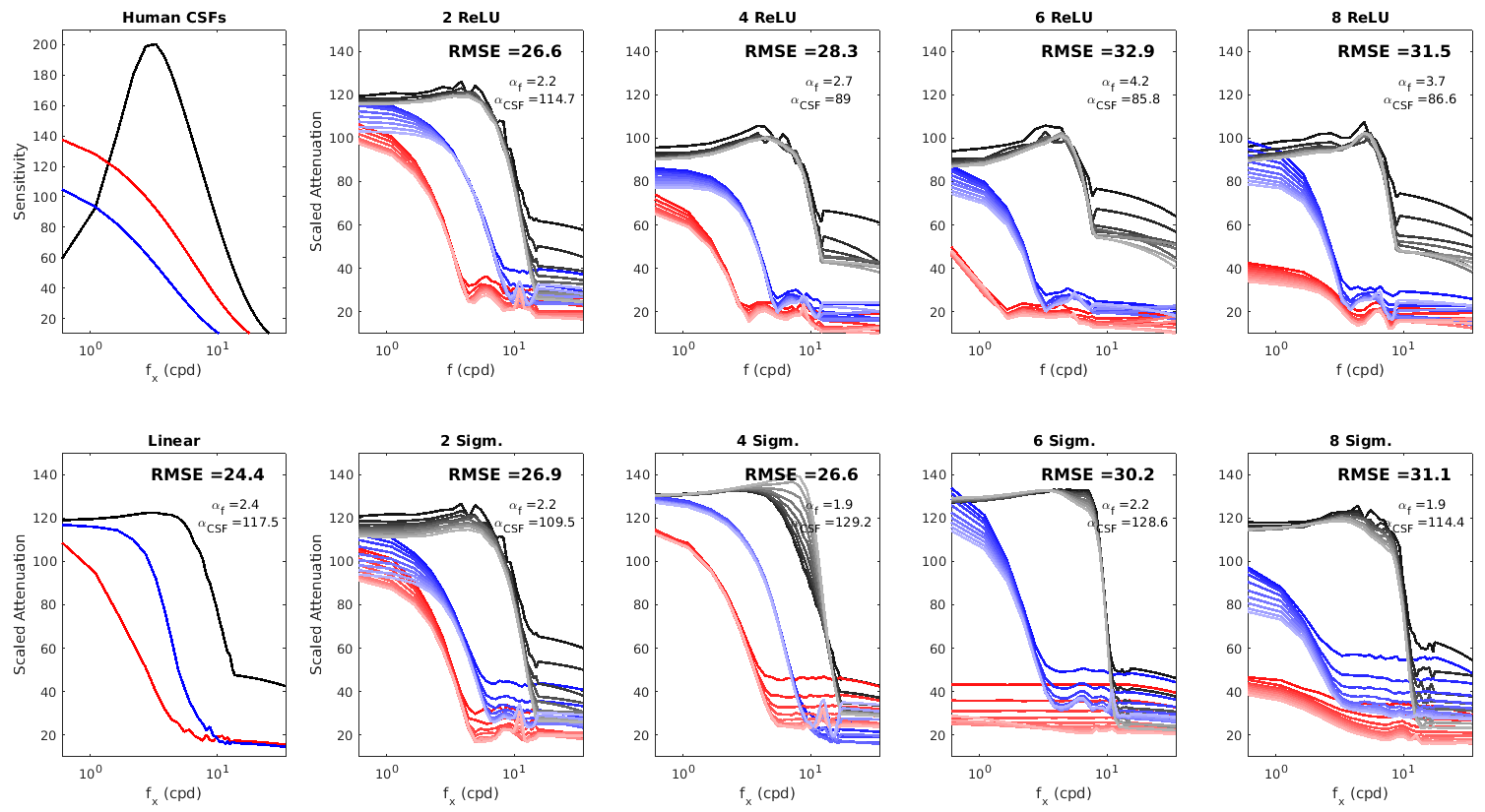}\\[-0.0cm]
   \caption{\textbf{Experiment 1 (Spatio-chromatic CSFs from the compensation of bio-distortion):} \blue{Attenuation factors for gratings of different contrast for a range of CNN autoencoders trained for bio-distortion compensation.
   Achromatic, blue and red lines refer to the CSFs of the achromatic, red-green, and blue-yellow channels respectively.
   Dark to light colors refer to progressively higher contrasts (evenly spaced in the range [0.07,0.6]). The human CSFs (top left) and the CSFs of the optimal linear solution (bottom left) are also shown as a convenient reference.
   RMSE measures describe the differences between the model and the human CSFs.
   The plots also display the values for the scaling factors in frequency and amplitude described in the text, Eq.~\ref{scaling}.}}
   \label{Exp1_csfs}
\end{figure}
\end{landscape}
}

\blue{Results show the emergence of a band-pass sensitivity in the achromatic channel and low-pass sensitivities in the chromatic channels.
The bandwidth of the chromatic channels is always substantially narrower than the achromatic bandwidth. These properties are \green{qualitatively} in line with human behavior.}

Shallower networks (either ReLU or Sigmoid) display bigger resemblance with human CSFs.
In particular, deeper nets introduce substantial distortion in the chromatic channels: note that the red-green channel is over attenuated (particularly for the 8-layer architectures but also in the 6-layer cases). The RMSE scores summarize these differences and show that shalower nets (2- and 4-layers) provide better explanations of the CSFs than deeper nets (6- and 8-layers).

\blue{Interestingly, the optimal linear solution (a single dense layer with identity activation) is the one that better reproduces the CSFs. However, by its linear nature, it cannot include contrast-dependent behavior.}
\blue{In this regard, the shallow networks (2-layers) display a consistent decay of the gain (attenuation factor) with contrast. This decay has an impact on the contrast response curves for gratings.
The contrast response curves describe the evolution of the amplitude of the response to a grating as a function of the contrast of the grating.
In humans contrast response curves are increasing saturating functions both for achromatic gratings~\cite{Legge80,Legge81} and for chromatic gratings~\cite{Uriegas97}.
The decay found in 2-layer CNNs implies a saturation of the contrast response curves for these shallow CNNs, in line with human behavior. Figure~\ref{nonlinear_resp_color} shows representative examples of these response curves: While the 2-layer network (top row) consistently displays saturating behavior for every frequency, the deeper net (bottom row) shows non-human (linear or expansion) responses.}

\blue{Finally, Figure~\ref{Exp1_csfs_lin} shows the CSFs corresponding to the linearized versions of the nonlinear CNNs, Eq.~\ref{global_linear_approx}.
Of course, the linear approximations have contrast-independent behavior and hence the same CSF for all contrasts.
The global linear approximations of the nonlinear models improve the resemblance of the CSFs with human behavior:
the linearized shallow nets are closer to humans than the linear model, and linearization corrects over attenuation of the chromatic channels in the 6-layer models.
However this increased similarity with human CSF comes at the cost of a significant drop in the performance (see the increase in $\varepsilon_{\textrm{LMS}}$ error in Table~1).
The linearization leads to rigid models which disregard the differences between the original nonlinear models and behave more similarly. In any case, linearization does not overcome the overattenuation of the red-green channel in the 8-layer models.}

\afterpage{
\begin{figure}[b!]
\begin{center}
\includegraphics[width=0.6\linewidth]{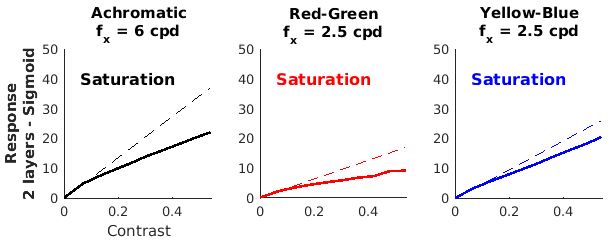}
\includegraphics[width=0.6\linewidth]{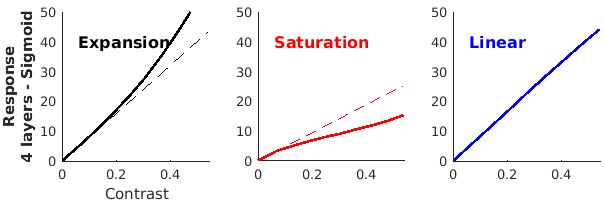}
   \caption{\textbf{Experiment 1 (Illustrative contrast responses from the compensation of bio-distortion):} \blue{Representative examples of nonlinear responses for achromatic and chromatic sinusoids found.
   Saturation in these responses comes from the decay in the attenuation factors with contrast in Fig.~\ref{Exp1_csfs}. Similarly, expansion comes from the increase in the attenuation factors.
   The linear behavior at the low-contrast regime has been plotted with dashed line as useful reference to highlight the nonlinear behavior. \green{It is interesting to note that the saturating or expansive nature of the final contrast nonlinearity is a \emph{collective} behavior that is not trivially attached to the specific (saturating or expansive) nonlinearities in individual neurons.}}
   }
   \label{nonlinear_resp_color}
\end{center}
\end{figure}
}

\afterpage{
\begin{landscape}
\vspace{5cm}
\begin{figure}[!h]
\vspace{4cm}
   \centering
    \includegraphics[width=1.22\textwidth]{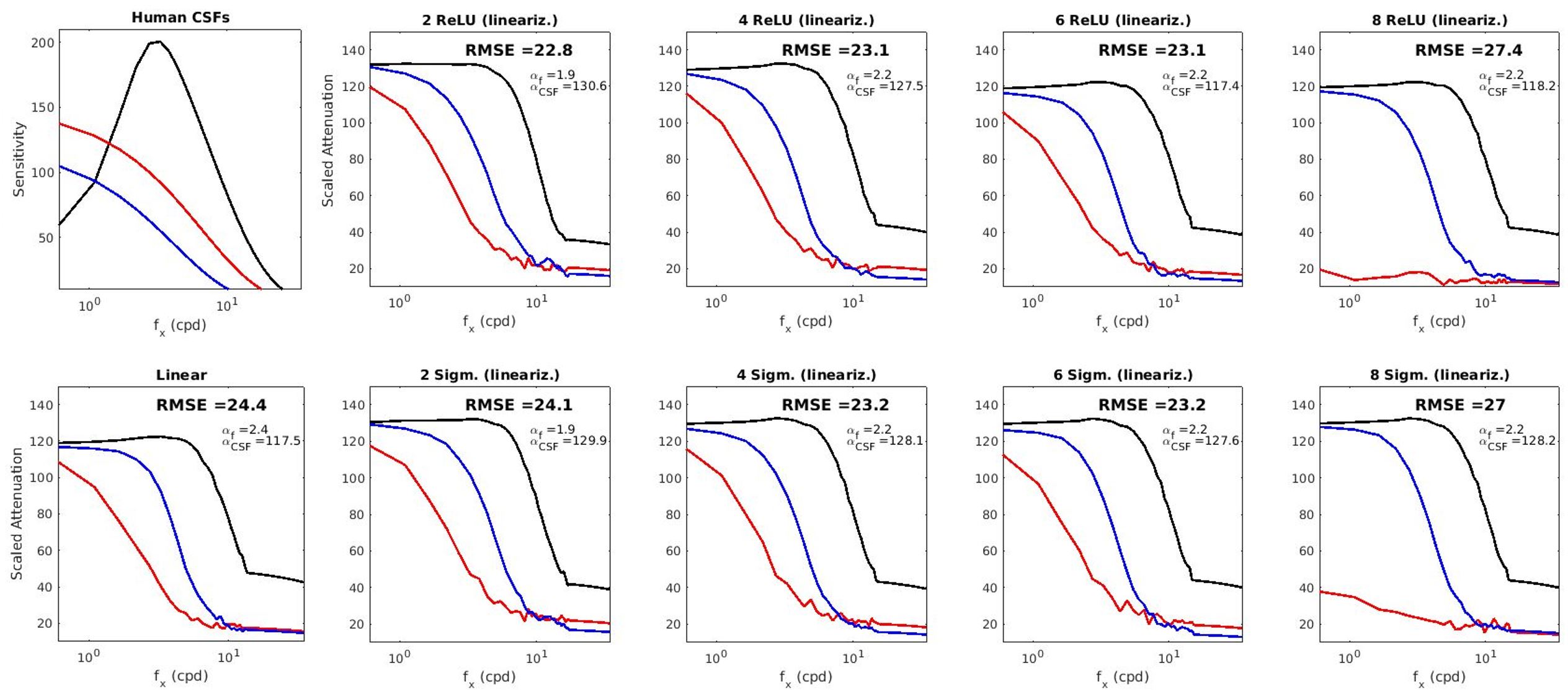}\\[-0.0cm]
   \caption{\textbf{Experiment 1 (Spatio-chromatic CSFs for linearized networks in bio-distortion compensation):} \blue{The human CSFs and the CSF of the optimal linear network are also included as reference.}
   }
   \label{Exp1_csfs_lin}
\end{figure}
\end{landscape}
}


\blue{Table~1 shows that while deeper networks are significantly better in fulfilling the computational goal (as expected from their increased flexibility), they are worse than shallow nets in reproducing the human behavior (as seen in Figs.~\ref{Exp1_csfs}-\ref{Exp1_csfs_lin}).}

\blue{Progressive improvement in the goal for increasing depth is numerically substantial (and also visible in the reconstructed signals in Fig.~\ref{Exp1_converg_perform} in Appendix A) from 2, 4, to 6 layers, and the numerical performance stays (statistically) the same for 8 layers. For this last case there are chromatic issues in line with what was been found in the CSFs:
the colorfulness of the reconstruction in Fig.~\ref{Exp1_converg_perform} is related to the relative gain of the chromatic channels. In particular, the consistent underestimation of the red-green CSF by the 8-layer CNNs (either using ReLU or Sigmoid activation) leads to a low-saturation images. Interestingly, this effect is also visible in the reconstructed images coming from the linearized CNNs, see Fig.~\ref{Exp1_visual_perform_lin}, and is consistent with the strong attenuation of the RG channel in the linearized 8-layer architectures in Fig.~\ref{Exp1_csfs_lin}.}

\blue{It is important to stress that the deviations in the chromatic CSFs in deep models do not come from not fulfilling the goal or having poor convergence in the
training. First, all models (even the linear one) do reduce the error of the original retinal degradation (see Table~1) so they are fulfilling the goal. And second, the learning curves in the Appendix~C (Figure~\ref{Exp1_converg_perform}) show that all models achieved a plateau in the training thus indicating proper convergence. Moreover, the asymptotic values achieved in the learning are consistent with the $\varepsilon_{\textrm{LMS}}$ in test shown in Table~1.}

\green{As stated above, RMSE errors in Table~1 have to be interpreted in terms of the scale of the human CSF. For example, the best and the worst CNNs (RMSE = 24.4 and RMSE = 33.1 respectively) have average deviations of 11\% and 15\% with regard to the maximum human sensitivity.
Of course, a single figure of merit averaged over frequencies and chromatic channels may hide an uneven distribution of the errors. For instance, consider the specific 6-layer-sigmoid CSF shown in Fig.~\ref{Exp1_csfs}, which displays a clear over attenuation of the red green channel.
In that case, if the global RMSE = 30.2 is broken down into its chromatic components we have 29.0, 35.0 and 26.8 for the achromatic, red-green and yellow-blue errors, which clearly point out that the biggest problem is in the red-green sensitivity. The same is true for the average over spatial frequencies: the global description does not stress the discrepancy in the low frequencies of the achromatic channel. That is why the (necessarily limited) description in the tables comes together with the explicit plots of the three CSFs for different contrasts.}

\begin{table}[t!]
\centering
\begin{tabular}{|c|cc|cc||cc|cc|}
\hline
                 &  \textbf{Comput. Goal}  &        &  \blue{\textbf{CSFs}}      &   &  \textbf{Comput. Goal}&  & \blue{\textbf{CSFs}}      & \\
                 &  $\varepsilon_{\textrm{LMS}}$            &        &  \blue{RMSE} &   &  $\varepsilon_{\textrm{LMS}}$          &                & \blue{RMSE}  & \\ \hline\hline
Bio-distortion      &  $15.5 \pm 0.2$         &        &  &                 &         & &  & \\ \hline\hline
Linear Net       &  $\mathbf{13.1} \pm 0.1$&        & $\blue{\mathbf{24.4}}$  &   &         & &  & \\ \hline\hline
CNNs             & Sigmoid   & ReLU        &  Sigmoid    &  ReLU      & Sigmoid     & ReLU       & Sigmoid     & ReLU       \\
                 & Nonlinear & Nonlinear   &  Nonlinear  &  Nonlinear & Linearized  & Linearized & Linearized  & Linearized \\ \hline
2-Layers         & $\mathbf{10.3}\pm 0.1$ & $\mathbf{10.8}\pm 0.1$  &  $\blue{\mathbf{26.9}}$  & $\green{\mathbf{24.5}\pm 0.3}$ & $12.5 \pm 0.1$ & $12.6 \pm 0.2$   &  \blue{24.1}   & \blue{22.8} \\ 
4-Layers         & $\mathbf{8.9} \pm 0.1$ & $\mathbf{9.1} \pm 0.1$  &  $\blue{\mathbf{26.6}}$  & $\blue{\mathbf{28.5}}$ & $12.5 \pm 0.1$ & $12.5 \pm 0.1$   &  \blue{23.2}   & \blue{23.1} \\ 
6-Layers         & $\mathbf{8.7} \pm 0.1$ & $\mathbf{8.5} \pm 0.1$  &  $\green{\mathbf{29.7}\pm 0.7}$  & $\blue{\mathbf{33.1}}$ & $12.5 \pm 0.2$ &  $12.5 \pm 0.2$   &  \blue{23.2}  & \blue{23.1} \\ 
8-Layers         & $\mathbf{8.9} \pm 0.2$ & $\mathbf{8.7} \pm 0.1$  &  $\blue{\mathbf{31.2}}$  & $\blue{\mathbf{31.6}}$ & $12.6 \pm 0.1$ & $12.7 \pm 0.1$  &  \blue{27.0}   & \blue{27.4} \\ \hline \hline
\end{tabular}
\label{table_RMSE_images}
\caption{\textbf{Experiment 1: Emergence of CSFs from bio-distortion compensation (computational goal and error in CSFs). } \blue{The achievement of the computational goal is described by $\varepsilon_{\textrm{LMS}}$ (error of the reconstructed signal in LMS) for batches of images of the independent test set (averages and standard deviations from 20 realizations with 50 images/batch).
The distance between the CSFs of the networks and the human CSFs is measured by the RMSE between the functions, Eq.~\ref{def_RMSE}. \green{Uncertainty of the RMSE was estimated only in two cases (2-layer ReLU and 6-layer sigmoid) and is represented here by the standard error of the corresponding means.} It is interesting to note that the optimal linear solution (computed from the train set) has worse performance in the test set than the linearized versions of the networks.}}
\end{table}

\green{Another important technical issue is the consistency of the CSFs over random initialization.
This is easy to check by training \emph{a number of times} the same architecture for the same computational task and over the same set of stimuli but from different initial values of the model parameters.
Given the intensive computation required\footnote{In our computer cluster typical training of the 2D models takes about 10 to 20 hours.} we checked this variability only in two illustrative models:
one with reasonably human-like behavior (2-layer ReLU), and another with less-human CSFs (6-layer sigmoid).
In these two models we re-trained them 20 additional times and recomputed the corresponding CSFs (results not plotted).
In the 6-layer case \emph{all} the explored seeds lead to a flat red-green CSF of too low sensitivity (i.e. a non-human behavior), and in some cases even the blue-yellow sensitivity was strongly attenuated too. On the contrary, the 2-layer case systematically leads to better CSFs, as summarized by the RMSE in Table~1, where the uncertainty is represented by the standard error of the mean. The shape of the sensitivities is pretty consistent in both cases, always better for the 2-layer case.
At the same time, and not surprisingly, the 6-layer architectures systematically led to lower $\varepsilon_{\textrm{LMS}}$ error.
Only 1 out of the 42 realizations (21 per model) led to a clear outlier (RMSE = 43.8 in the 6-layer case) and even for this CSF-outlier the distortion $\varepsilon_{\textrm{LMS}}$ was not off the distribution.
According to the observed consistency, the remaining 49 configurations of task/architecture in the work were studied using a single random initialization of the parameters.
}

\blue{The next experiments explore the consistency of the human-like behavior found in shallow autoencoders
in a number scenarios.
According to the results found in Experiment 1, we select the 2-layers-ReLU autoencoder as a representative example of shallow architecture with reasonable human-like behavior (RMSE of 11\% of the maximum sensitivity) so we focus on this architecture in Experiments 2-4.}


\subsection{Experiment 2: Consistency of CSFs over a range of bio-distortions}

\begin{figure}[t!]
\begin{center}
\begin{centering}
\hspace{-0cm}\includegraphics[width=0.77\linewidth]{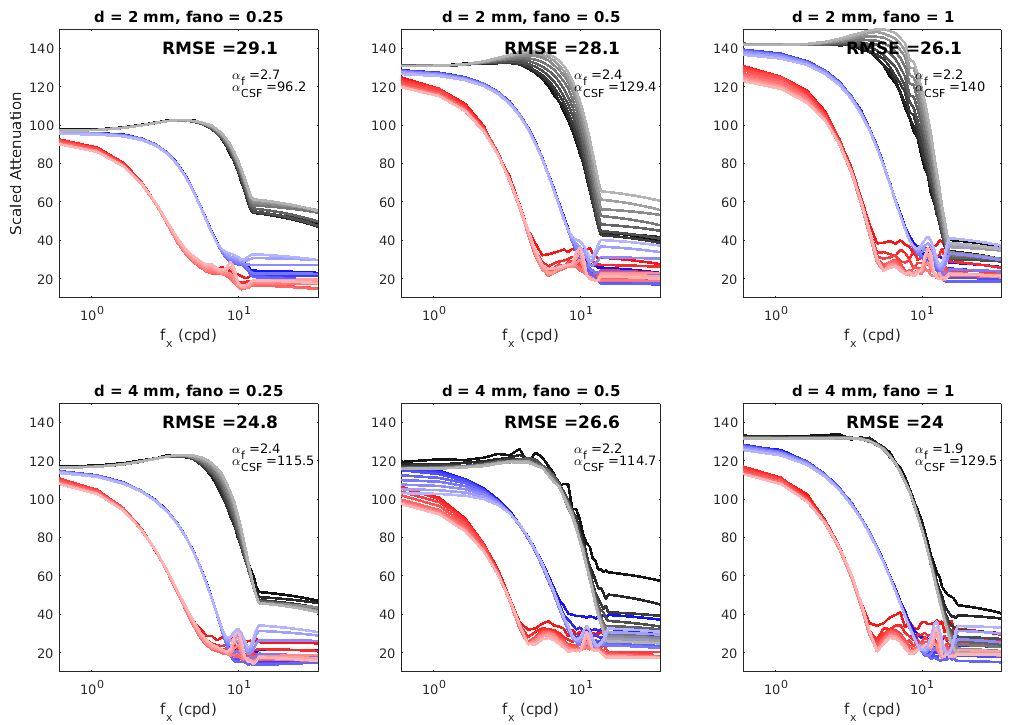}\\
\hspace{0.43cm}\includegraphics[width=0.74\linewidth]{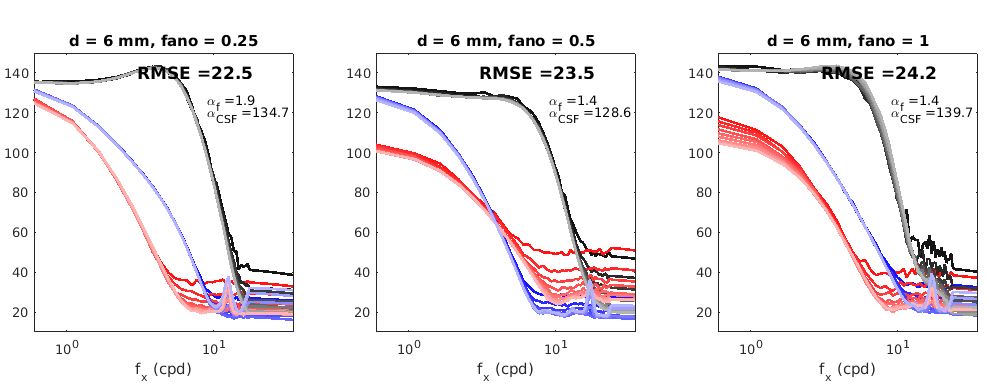}
\end{centering}
   \caption{\textbf{Experiment 2: Consistency of human-like result over a range of retinal degradations.}
   \blue{Spatio-chromatic CSFs of the 2-layer ReLU net for a range of retinal degradations. The RMSE distortion over the nine retinal conditions is $25 \pm 2$ (mean and standard deviation). The line style conventions and meaning of the numbers is the same as in Experiment 1.}}
   \label{Exp2_csfs}
\end{center}
\end{figure}

\blue{Figure~\ref{Exp2_csfs} shows the CSFs obtained when training the 2-layer-ReLU net to compensate a range of retinal degradations (as described by the different pupil diameters and Fano factors).
Learning curves that show the good convergence of the models and representative visual examples of the reconstructions are given in the Appendix~C (Figures~\ref{Exp2_converg} and~\ref{Exp2_performance}).}

\blue{Results in Fig.~\ref{Exp2_csfs} show that band pass / low-pass channels with distinct bandwidths consistently appear in all cases, and the RMSE with human CSF ($25 \pm 2$), mean and standard deviation, stays in the low range of the values found in Experiment 1.}

\blue{Regarding the evolution of the CSFs with contrast, it is important to note that for some conditions (low blur and high noise) the gain in the achromatic CSF increases with contrast, which is equivalent to contrast response curves that are not saturating.}

\subsection{Experiment 3: CSFs from chromatic adaptation versus bio-distortion compensation}

\blue{Figure~\ref{Exp3_4_csfs} (top row) shows the CSFs emerging when the representative shallow network with human-like behavior in Experiment 1 is trained for a range of alternative low-level tasks, some involving compensation of the retinal degradation (1st, 2nd, and 3rd cases), and some others only involving chromatic adaptation (4th and 5th). The corresponding learning curves for the models and visual examples of reconstruction are given in the Appendix~A (Figure~\ref{Exp3_converg_performance}).
Table 2 (top panel) summarizes the numerical performance of the models in this experiment ($\varepsilon_{\textrm{LMS}}$ for the computational goal, and RMSE for the CSFs).}

\blue{First, lets focus on the case where the determination of the CSF faithfully follows Eq.~\ref{csf1} and hence we have a realistic eye+retina degradation (solid lines). Results show that only the cases where the task involves the bio-degradation imply a clear difference in bandwidth between the achromatic and the chromatic channels.
In the cases where there is only chromatic adaptation, the three CSFs are wider and of similar bandwidth. This behavior is clearly non-human, as confirmed by the RMSE measures in the 4th and 5th panels at the right.}

\blue{Second, this difference is more clear in the idealized cases, Eq.~\ref{csf_net}, where clean sinusoids are used to determine the CSFs (dashed-lines). In this situation the CSFs of the purely chromatic goals are wider and flatter indicating that the networks are not performing any particular spatial modification in any chromatic channel. As a result, the RMSE values for the chromatic adaptation cases (light style numbers below the frequency axis) substantially increase indicating poorer description of human CSFs. In this regard, the errors for the cases in which the task involves bio-degradation are lower, but they are even lower if the CSF is measured considering the realistic degradation in the input.}

In summary, the results show two trends. On the one hand, human-like features emerge in the CSFs if the degradation-compensation task is considered, but they do not if only chromatic adaptation is considered. On the other hand, the CSFs are closer to human \green{in RMSE} if the determination takes the retinal degradation into account in the sinusoids.

\blue{Finally, there is an interesting human chromatic feature that is well captured by all the CNN models that were trained for chromatic adaptation: all of them display a sort of Von-Kries modification of the RG and YB channels. Note that when the red illuminant has to be compensated (3rd and 5th cases), the red-green CSF is attenuated while the blue-yellow CSF is boosted, and the other way around in in the compensation of a bluish illuminant (2nd and 4th cases, where the blue-yellow channel is attenuated).}

\subsection{Experiment 4: Consistency of spatio-chromatic CSFs under changes of signal statistics}

\blue{Figure~\ref{Exp3_4_csfs} (bottom row) shows the CSFs emerging when the representative shallow network with human-like behavior in Experiment~1 is trained for the range of low-level tasks considered in Experiment~3 optimizing the performances over cartoon images (as opposed to regular photographic images). The corresponding learning curves for the models and visual examples of reconstruction are given in the Appendix~C (Figure~\ref{Exp4_converg_performance}).
Table 2 (bottom panel) summarizes the numerical performance of the models in this experiment.}

\afterpage{
\begin{landscape}
\vspace{3cm}
\begin{figure}[!h]
\vspace{2cm}
   \centering
   \hspace{-0cm}\includegraphics[width=1.22\textwidth]{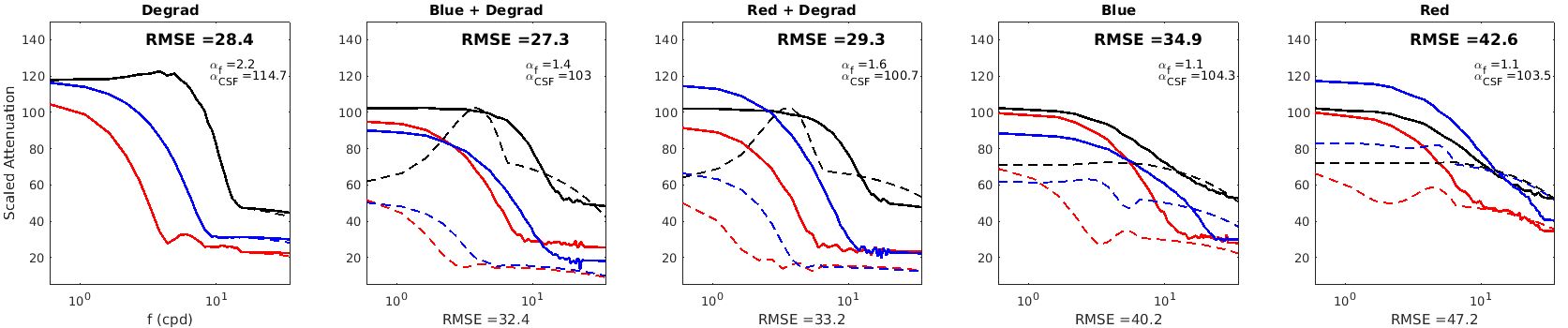}\\[1cm]
   \hspace{-0cm}\includegraphics[width=1.22\textwidth]{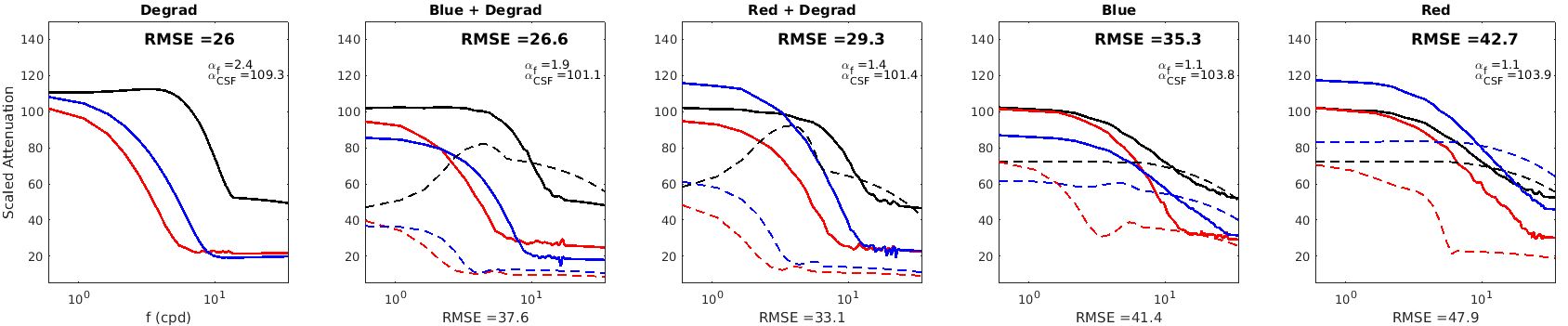}\\
    \caption{\green{\textbf{Experiment 3 (top row): CSFs from bio-distortion compensation versus chromatic adaptation in natural images.}
    From left to right we consider the CSFs that emerge from (a) the compensation of bio-degradation (left plot), (b) combinations of chromatic adaptation and degradation compensation (2nd and 3rd plots), and (c) pure chromatic adaptation panels (4th and 5th plots at the right).
    The cases in the columns 1st-to-3rd involve retinal degradation (with or without color adaptation) while the cases in the 4th-5th columns only involve chromatic adaptation.
    For the sake of clarity, only low contrast results are shown.
    Solid lines correspond to CSFs determined using Eq.~\ref{csf1} (where the input to the network includes realistic acquisition process). Dashed lines correspond to CSFs determined using Eq.~\ref{csf_net} (where the network is probed with clean sinusoids).
    The RMSE values in bold correspond to the CSFs determined in realistic conditions (curves in bold). The RMSE values displayed below the frequency axis correspond to the CSFs determined with clean sinusoids (dashed lines).
    In the cases involving chromatic adaptation the input sinusoids were color shifted according to the corresponding change in the illuminant.}\\
    \green{\textbf{Experiment 4 (bottom row): Consistency of the results for different signal statistics (cartoon images).} \green{The computational goals are the same as above. The only difference is that the models are trained with cartoon images (from the Pink Panther Show) as opposed to regular photographic images from ImageNet (used in Experiments 1-3).}}}
    \label{Exp3_4_csfs}
\end{figure}
\end{landscape}
}

\afterpage{
\begin{table}[t!]
\centering
\begin{tabular}{|c|c|c|c||c|c|}
\hline
\textbf{NATURAL Images}   &                           &        &      &      &            \\ \hline\hline
\textbf{Comput. Goals}    &  \blue{\textbf{Degradat.} Only}    &  \blue{\textbf{Degradat.} + Blue Adapt.}  & \blue{\textbf{Degradat.} + Red Adapt.}  & \blue{Blue \textbf{Adapt.}}      &  \blue{Red \textbf{Adapt.}} \\ \hline
Original $\varepsilon_{\textrm{LMS}}$              &  $12.1 \pm 0.2$            &   $18.1 \pm 0.2$                   &   $15.2 \pm 0.3$              &   $13.2 \pm 0.2$        &  $10.2 \pm 0.3$   \\
$\varepsilon_{\textrm{LMS}}$ after 2-layers ReLU             &   $9.79 \pm 0.08$           &   $8.7 \pm 0.1$                   &  $8.9 \pm 0.1$                &   $1.92 \pm 0.05$       & $1.01 \pm 0.01$    \\  \hline
\textbf{RMSE of CSFs}   &                           &        &      &      &            \\ \hline
$\textbf{CSF}_{\textrm{realist.}}$    &  \blue{$\mathbf{28.4}$}    &   \blue{$\mathbf{27.3}$}           &   \blue{$\mathbf{29.3}$}       &   \blue{$\mathbf{34.9}$}&  \blue{$\mathbf{42.6}$}   \\
${CSF}_{\textrm{simplist.}}$   &                            &  \blue{${32.4}$}    &   \blue{${33.2}$}           &   \blue{${40.2}$}       &   \blue{${47.2}$}   \\ \hline\hline
\textbf{CARTOON Images}   &                           &        &      &      &            \\ \hline\hline
\textbf{Comput. Goals}     &  \blue{\textbf{Degradat.} Only}    &  \blue{\textbf{Degradat.} + Blue Adapt.}  & \blue{\textbf{Degradat.} + Red Adapt.}  & \blue{Blue \textbf{Adapt.}}      &  \blue{Red \textbf{Adapt.}} \\ \hline
Original $\varepsilon_{\textrm{LMS}}$               &  $12.9 \pm 0.2$            &   $18.1 \pm 0.1$                   &   $15.6 \pm 0.2$              &   $13.4 \pm 0.3$        &  $9.1 \pm 0.3$   \\
$\varepsilon_{\textrm{LMS}}$ after 2-layers ReLU             &   $10.14 \pm 0.08$           &   $9.5 \pm 0.1$                   &  $9.2 \pm 0.1$                &   $1.34 \pm 0.02$       & $0.837 \pm 0.06$    \\  \hline
\textbf{RMSE of CSFs}   &                           &        &      &      &            \\ \hline
$\textbf{CSF}_{\textrm{realist.}}$   &  \blue{$\mathbf{26.0}$}    &   \blue{$\mathbf{26.6}$}           &   \blue{$\mathbf{29.3}$}       &   \blue{$\mathbf{35.3}$}&  \blue{$\mathbf{42.7}$}   \\
${CSF}_{\textrm{simplist.}}$  &     &  \blue{${37.7}$}    &   \blue{${33.1}$}           &   \blue{${41.4}$}       &   \blue{${47.9}$}    \\ \hline\hline
\end{tabular}
\label{table_RMSE}
\caption{\textbf{Experiment 3 (top panel): Compensation of bio-degradation versus Chromatic adaptation in natural images}. \blue{Performance in the goals and eventual human-like CSFs
are described by $\varepsilon_{\textrm{LMS}}$ and RMSE respectively. CSF$_{\textrm{realist.}}$ and CSF$_{\textrm{simplist.}}$ refer to the way the CSF is computed (taking into account or neglecting the retinal degradation in the sinusoidal stimuli)}.
\textbf{Experiment 4 (bottom panel): Consistency under change of image statistics.} \blue{The considered goals and magnitudes have the same meaning as in Exp. 3. The only difference is in the scenes used to train the models.}
}
\end{table}
}

The parallelism in the results of Experiments 3 and 4 confirms the robustness of the behaviors shown in Experiment 3
to certain changes in signal statistics. Note that this parallelism doesn't mean that the CSFs are independent of the signal statistics.
It just means that they are invariant to \emph{this} change of statistics.
It is important to remark that the low-level statistics of these (apparently different) sources may not be that different.
Colors of the Pink Panther images are certainly more saturated, but beyond this obvious fact, other differences may be subtle.
In particular, we took precautions to get frames from a 5 hours compilation where backgrounds around the whole chromaticity range
appear not to bias the chromatic CSFs.
Regarding the spatial content, note that there are plenty of edges of arbitrary orientations and also low-frequency
transitions and shadows in the cartoons. More radical modifications of spatial information (e.g. edit the cartoon images to make them
isoluminant -i.e. zero contrast in the achromatic channel-) could lead to substantial variation of the CSFs, but the goal of this
illustration is to point out the robustness of the result more than look for its limits.

\subsection{\green{Experiment 5:  CSFs from bottleneck compensation versus bio-distortion compensation}}

\green{Figure~\ref{Exp5A_csfs} shows the CSFs of the systems that emerge from the architectures with bottlenecks considered in Fig.~\ref{Archi}-right, when considering two different functional goals: (1)~\emph{pure reconstruction} of the input signal, i.e. compensation of the information loss imposed by the bottleneck, and
(2)~\emph{compensation of the bottleneck together with compensation of the bio-distortion}. Table~\ref{Exp5A_table} summarizes the distortions in the CSFs, RMSE, and the performance in the reconstruction, $\varepsilon_{\textrm{LMS}}$.}

\green{The Appendix C, Fig.~\ref{Exp5A_converg},  confirms that these architectures converged to a plateau of $\varepsilon_{\textrm{LMS}}$. Moreover,  consistently with the data in Table~\ref{Exp5A_table}, Figs.~\ref{Exp5A_converg} and~\ref{Exp5A_visual_perform} show that these systems achieve the computational goals to an extent that depends on the severity of the bottleneck in a very intuitive fashion (see comments in Appendix C).}

\green{More interesting is what happens to the emerging CSFs in Fig.~\ref{Exp5A_csfs}.
In the absence of a bottleneck, pure reconstruction leads to wide filters equal in the three chromatic channels, a clearly non-human result with RMSE~$\sim 40$ (architectures A and B). Similarly to pure chromatic adaptation, unconstrained pure reconstruction induces no spatial selectivity and hence small similarity with human vision. Mild bottlenecks restricting the number of features and/or the spatial resolution do introduce differences in the bandwidth of the achromatic/chromatic channels, but the shape of the filters is far from human ($\textrm{RMSE} \geq 30$ in architectures C - D). Then, more severe bottlenecks (architectures E-G and U-Net) quickly leads to over-attenuation of one or both chromatic CSFs (and hence non-human behavior with RMSE~$\sim 35$ in these architectures for reconstruction).
On the other hand, the very same architectures trained for the compensation of bio-distortion lead to more human-like CSFs. See the band-pass / low-pass shape of the achromatic/chromatic CSFs and the RMSE~$\sim 25$ except for architectures F and G that over attenuate the chromatic CSFs but still preserve the band-pass nature of the achromatic channel.}
\green{Better preservation of chromatic CSFs by the systems tuned to compensate the bio-distortion is visually confirmed by the reconstructions of a representative image in Appendix C, Fig.~\ref{Exp5A_visual_perform}.}

\green{In summary, pure reconstruction with the explored bottlenecks induces a difference between the relative bandwidths of the achromatic and chromatic CSFs. However, the results become closer to human (both in the shape of the filters and in RMSE) when considering the compensation of the bio-degradation of the retinal signal. And this resemblance remains even if the system is not constrained by a bottleneck.}

\afterpage{
\begin{landscape}
\vspace{3cm}
\begin{figure}[!h]
\vspace{2cm}
   \centering
   \hspace{-0cm}\includegraphics[width=1.22\textwidth]{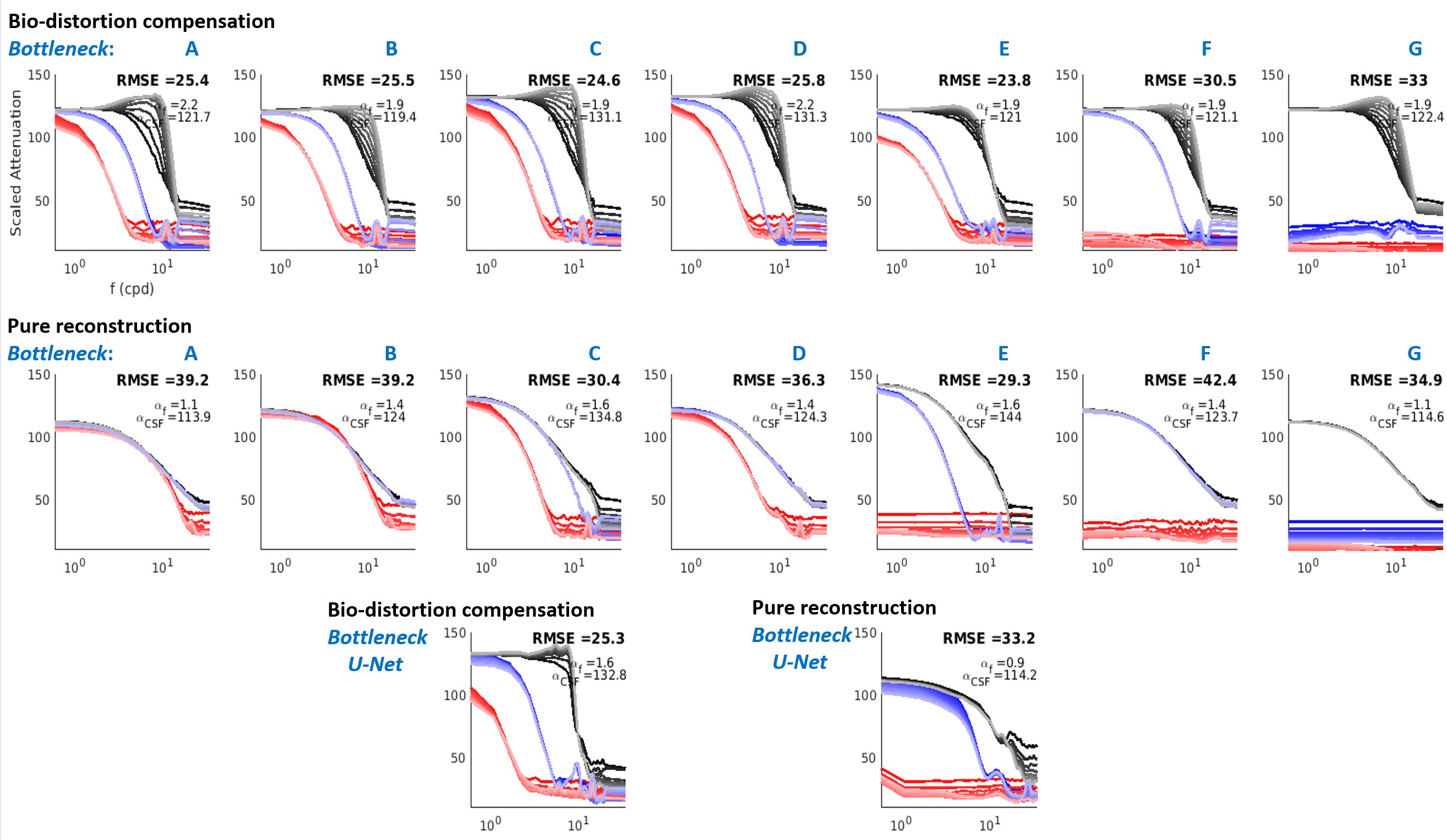}\\
    \caption{\textbf{Experiment 5: CSFs from architectures with bottlenecks} Notation of the architectures with bottlenecks (letters in blue) refers to the diagrams shown in Fig.~\ref{Archi}.
    \textbf{(Top row):} CSFs emerging from bottleneck compensation and bio-distortion degradation in progressively more restrictive bottlenecks in a 4-later architecture.
    \textbf{(Middle row):} CSFs emerging from pure reconstruction of the signal in the same architectures.
    \textbf{(Bottom row):} CSFs emerging in U-Nets from bio-distortion compensation (left) or pure reconstruction (right).
    }
   \label{Exp5A_csfs}
\end{figure}
\end{landscape}
}


\begin{table}[]
\centering
\begin{tabular}{ll|cc|cccccc|}
\cline{3-10}
 &  & \multicolumn{2}{c|}{\textbf{No Bottleneck}} & \multicolumn{6}{c|}{\textbf{Bottlenecks}} \\ \cline{3-10}
 &  & \multicolumn{1}{c|}{\textbf{A}} & \textbf{B} & \multicolumn{1}{c|}{\textbf{C}} & \multicolumn{1}{c|}{\textbf{D}} & \multicolumn{1}{c|}{\textbf{E}} & \multicolumn{1}{c|}{\textbf{F}} & \multicolumn{1}{c|}{\textbf{G}} & \textbf{U-Net} \\ \hline\hline
\multicolumn{1}{|c|}{\multirow{2}{*}{\textbf{Bio-Distort.} }} & $\varepsilon_{\textrm{LMS}}$ & \multicolumn{1}{c|}{$9.1\pm0.3$} & $9.2\pm0.2$ & \multicolumn{1}{c|}{$9.6\pm0.1$} & \multicolumn{1}{c|}{$9.8\pm0.2$} & \multicolumn{1}{c|}{$10.3\pm0.4$} & \multicolumn{1}{c|}{$11.2\pm0.2$} & \multicolumn{1}{c|}{$15.6\pm0.5$} & $12.3\pm0.3$ \\ \cline{2-10}
\multicolumn{1}{|c|}{} & \textit{RMSE} & \multicolumn{1}{c|}{{\bf 25.4}} & {\bf 25.5} & \multicolumn{1}{c|}{{\bf 24.6}} & \multicolumn{1}{c|}{{\bf 25.8}} & \multicolumn{1}{c|}{{\bf 23.8 }} & \multicolumn{1}{c|}{{\bf 30.5}} & \multicolumn{1}{c|}{{\bf 33.0}} & {\bf 25.3} \\ \hline\hline
\multicolumn{1}{|c|}{\multirow{2}{*}{\textbf{Pure Recons.}}} & $\varepsilon_{\textrm{LMS}}$ & \multicolumn{1}{c|}{$1.2\pm0.1$} & $2.2\pm0.1$ & \multicolumn{1}{c|}{$4.8\pm0.3$} & \multicolumn{1}{c|}{$3.0\pm0.5$} & \multicolumn{1}{c|}{$12.8\pm0.7$} & \multicolumn{1}{c|}{$5.6\pm0.8$} & \multicolumn{1}{c|}{$11.9\pm0.5$} & $12.2\pm0.6$ \\ \cline{2-10}
\multicolumn{1}{|c|}{} & \textit{RMSE} & \multicolumn{1}{c|}{{\bf 39.2}} & {\bf 39.2} & \multicolumn{1}{c|}{{\bf 30.4}} & \multicolumn{1}{c|}{{\bf 36.3}} & \multicolumn{1}{c|}{{\bf 29.3}} & \multicolumn{1}{c|}{{\bf 42.4}} & \multicolumn{1}{c|}{{\bf 34.9}} & {\bf 33.2} \\ \hline
\end{tabular}
\caption{\green{\textbf{Experiment 5: Compensation of bio-distortion versus pure reconstruction in networks with bottlenecks.}
The bottlenecks in the architectures A-G and U-Net are described in Fig.~\ref{Archi}-right. Performance in the reconstruction goals is measured by the $\varepsilon_{\textrm{LMS}}$ error (in a test set) and the quality of the CSFs is given by the RMSE error. The considered bio-distortion was the central case in Fig.~\ref{fig_goals} with an original level of $\varepsilon_{\textrm{LMS}} = 15.5$, which is reduced to the values reported in the first row after the application of the models. In the pure reconstruction case the original retinal distortion is $\varepsilon_{\textrm{LMS}} = 0$ so the errors reported in the 3rd row come from poor reconstruction or an incomplete compensation of the bottleneck.}
}
\label{Exp5A_table}
\end{table}

\afterpage{
\begin{figure*}[h!]
\begin{center}
\begin{centering}
\hspace{0.2cm}\includegraphics[width=0.64\linewidth]{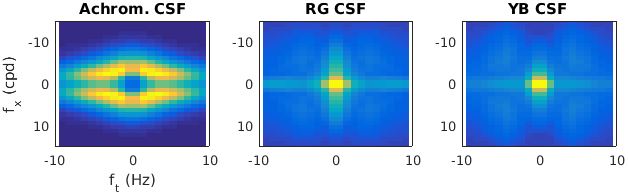}\\
\hspace{-0cm}\includegraphics[width=0.65\linewidth]{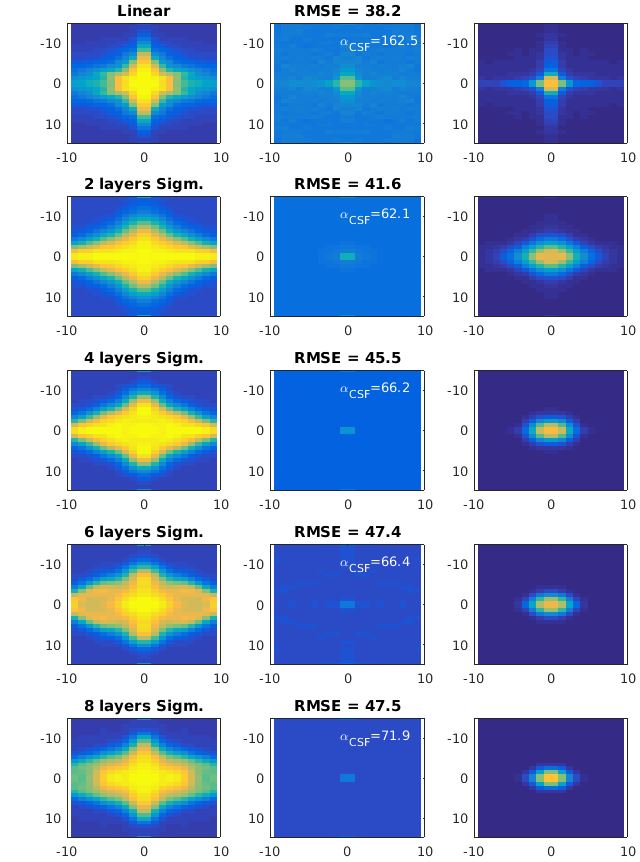}\\
\end{centering}
   \caption{\textbf{Experiment 6: Spatio-temporal-chromatic CSFs from bio-distortion compensation.} \blue{The first row shows the human CSFs in $(f_x,f_t)$, and the following show the model CSFs. The RMSE numbers (average over channels) represent the distance between them. In this experiment $\alpha_f = 1$ in all cases.}}
   \label{Exp5_csfs}
\end{center}
\end{figure*}
}

\afterpage{
\begin{figure*}[t!]
\begin{center}
\includegraphics[width=0.65\linewidth]{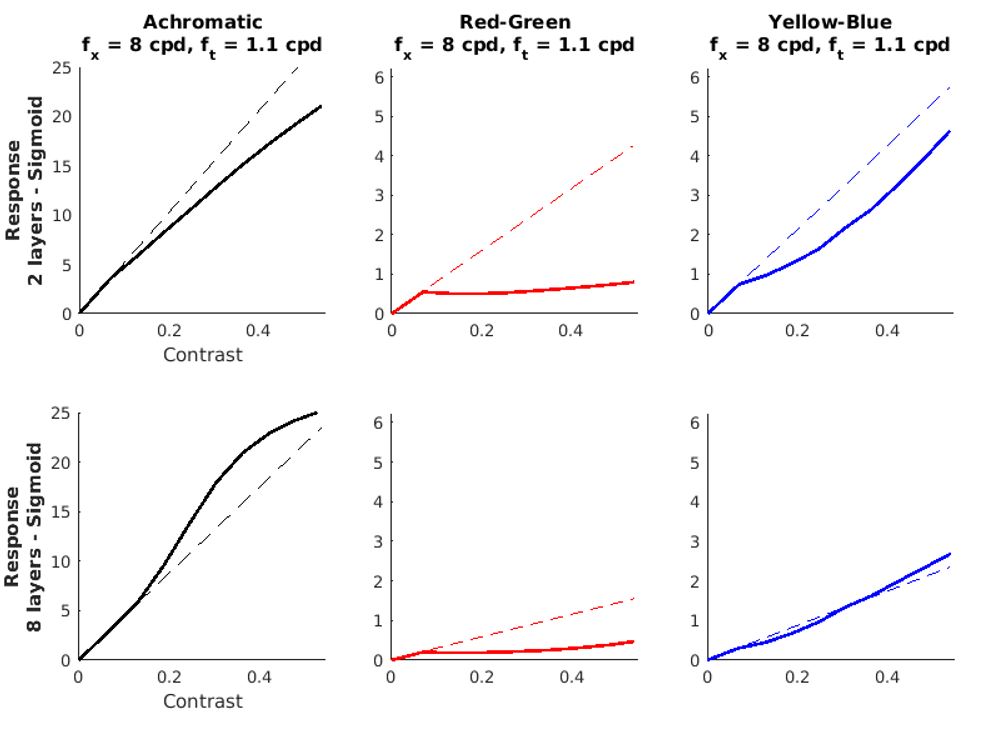}
   \caption{\textbf{Experiment 6:} Representative examples of nonlinear responses for spatio-temporal achromatic and chromatic gratings in shallow (top) and deeper (bottom) networks for bio-distortion compensation.}
   \label{nonlinearities_spatio_temp}
\end{center}
\end{figure*}
}

\subsection{Experiment 6: Spatio-chromatic-temporal CSFs from bio-distortion compensation}

\blue{Figure~\ref{Exp5_csfs} shows the attenuation factors found for low-contrast moving sinusoids (both achromatic and chromatic) in the plane $(f_x,f_t)$ for a range of 3D CNN autoencoders and for the optimal linear solution.
Experimental human CSFs for achromatic moving gratings~\cite{Kelly79}, and for chromatic moving gratings~\cite{Mara11} are also included as useful reference.
The learning curves for the models and visual examples of reconstructions are given in the Appendix~C (Figure~\ref{Exp5_converg_perform}).
Table 3 summarizes the numerical performance of the models in this experiment.
}

\afterpage{
\begin{table}[t!]
\centering
\begin{tabular}{|c|c|c|}
\hline
                 &  \textbf{Comput. Goal}  &  \blue{\textbf{CSFs}} \\
                 &  $\varepsilon_{\textrm{LMS}}$    &  \blue{RMSE}  \\ \hline\hline
Distortion       &  $5.2 \pm 0.1$         &   \\ \hline\hline
Linear Net       &  $\mathbf{3.5} \pm 0.2$& $\blue{\mathbf{38.2}}$   \\ \hline\hline
CNNs             & Sigmoid   & Sigmoid     \\
                 & Nonlinear & Nonlinear   \\ \hline
2-Layers         & $\mathbf{2.07} \pm 0.06$ &   $\blue{\mathbf{41.6}}$   \\ 
4-Layers         & $\mathbf{1.96} \pm 0.07$ &   $\blue{\mathbf{45.5}}$   \\ 
6-Layers         & $\mathbf{1.83} \pm 0.06$ &   $\blue{\mathbf{47.4}}$   \\ 
8-Layers         & $\mathbf{1.89} \pm 0.07$ &   $\blue{\mathbf{47.5}}$   \\ \hline \hline
\end{tabular}
\label{table_RMSE_video}
\caption{\textbf{Experiment 6: Emergence of spatio-temporal-chromatic CSF in 3D CNNs for compensation of bio-distortion.} \blue{The measures of the achievement of the goal $\varepsilon_{\textrm{LMS}}$ and the distance with human behavior (RMSE difference between model and human CSF) have the same meaning as in the rest of Experiments.
The degradation included in the movies was the same as the considered for images (same pupil diameter and Fano factor). However, the numerical $\varepsilon_{\textrm{LMS}}$ deviations turned out to be lower because the considered movies are \emph{darker} and hence smaller LMS values lead to substantially smaller Poisson noise.}.}
\end{table}
}

The CSF results show that the main feature of the spatio-temporal human window of visibility (its diamond shape), with smaller spatial bandwidth for higher temporal frequencies (or speeds)~\cite{Kelly79,Watson13} is reproduced by all the models as well as the substantially lower bandwidth of the chromatic channels, focused on very-low spatio-temporal frequencies. \green{The error of the best net is RMSE $=17\%$ of the maximum sensitivity.}

Consistently with the results found in images (Experiment 1), resemblance with human CSFs is bigger in shallower models (linear, 2-layers \green{with RMSE about 17\% or 18\% respectively}) than in deeper models (6-layers, 8-layers \green{with RMSE about 22\%}) despite the performance of the deeper models in the goal is substantially better than the performance of the linear or the 2-layer model.
The major differences are in the scaling of the chromatic CSFs: note that deeper models over attenuate the chromatic patterns.
The RMSE measures confirm the superiority of the shallower solutions.
For instance, note that the over attenuation of the red-green channel in the CNNs implies that the greenish hue of the background in the visual example of Fig.~\ref{Exp5_converg_perform} fades away, while it does not in the linear solution (which has obvious problems in other respects).

\blue{The linear solution cannot display a contrast dependent behavior, but the 2-layer architecture displays a consistent decay of the gain with contrast that is in line with the saturating nature of contrast response curves of humans for moving sinusoids~\cite{Simoncelli98,Morgan06}.
Figure~\ref{nonlinearities_spatio_temp} shows illustrative examples of these response curves:  While the 2-layer network (top row) consistently displays saturating behavior, the deeper net (bottom row) shows bigger variability on the shape of the response.}

\blue{As in the image case, the deviations in the chromatic CSFs in deep models do not come from not fulfilling the goal or having poor convergence in the training. First, all models (even the linear one) do reduce the error of the original retinal degradation so they are solving the computational problem.  And second, the learning curves in the Appendix~C (Figure~\ref{Exp5_converg_perform}) show that all models achieved a plateau in the training thus indicating proper convergence. Moreover, the asymptotic values achieved in the learning are consistent with the $\varepsilon_{\textrm{LMS}}$ in test shown in Table 4.}



\vspace{0.4cm}
\section{4. Discussion}

\subsection{Summary of results}

\blue{In the experiments we trained a range of CNN autoencoders over natural scenes to
solve different low-level vision goals: the compensation of retinal distortions,
the compensation of changes in the illumination, \green{the compensation of information loss after simple bottlenecks (or pure reconstruction after bottlenecks)}, and combinations of these.}


\green{Following the analysis of linearized networks presented  in~\cite{Gomez20b} it makes sense to stimulate these nets with achromatic, red-green and yellow-blue isoluminant sinusoids and moving sinusoids.}
The attenuation suffered by these gratings shows that:
\begin{itemize}

\item Human-like CSFs may emerge in systems that compensate retinal distortion: specifically, 2D~shallow autoencoders trained to compensate retinal distortion display narrow low-pass behavior in the chromatic channels and wider band-pass behavior in the achromatic channel, so the shape and relative bandwidth of these artificial CSFs resemble those of humans (Figs.~\ref{Exp1_csfs},    \ref{Exp1_csfs_lin}, and~\ref{Exp2_csfs}). \green{Of course the match is not complete:
the best CSFs obtained from the explored CNNs still deviate from
human CSFs (RMSE~$\sim 11\%$ of the maximum sensitivity).}
Deeper autoencoders for the same goal also show CSFs with these basic shapes but the resemblance with human CSFs is consistently lower \green{(RMSE~$\sim 15\%$ of the maximum sensitivity)}, particularly due to poor scaling of the chromatic CSFs (Figs.~\ref{Exp1_csfs} and~\ref{Exp1_csfs_lin}).

\item \green{Artificial CSFs obtained from the compensation of retinal distortion differ from human CSFs in two qualitative aspects:
(a) The decay of network sensitivity found at low frequencies for achromatic gratings is not as big as in humans, and (b) The relative amplitude of the red-green and the yellow-blue CSFs in the networks is inverted with regard to the humans. In our networks the YB CSF is always bigger than the RG CSF, and interestingly, this is pretty consistent over different architectures and datasets with different image statistics.}

\item Similar sensitivities consistently appear in shallow autoencoders for a range of levels in retinal distortions (Fig.~\ref{Exp2_csfs}).

\item Human-like CSFs with distinct bandwidths in achromatic/chromatic channels do not appear in pure chromatic adaptation tasks, but they do as soon as the retinal distortion compensation goal is considered (with or without chromatic adaptation). \green{The compensation of chromatic shifts together with the compensation of bio-distortion leads to systems in which the chromatic CSFs change their global gain similarly to a Von-Kries mechanism (Fig.~\ref{Exp3_4_csfs}, top).}

\item CSFs emerging from chromatic adaptation and degradation compensation goals are similar for natural images and cartoon images (Fig.~\ref{Exp3_4_csfs}, bottom).

\item  \green{Pure reconstruction in architectures with a restrictive bottleneck induces changes in the relative bandwidths of the achromatic and chromatic CSFs with regard to trivial all-pass filters found in systems without bottleneck. However (in the explored cases) these CSFs are remarkably non-human. Interestingly, the very same architectures lead to more human-like CSFs as soon as the retinal distortion compensation goal is considered (Fig.~\ref{Exp5A_csfs}).}

\item 3D autoencoders for retinal degradation compensation display a wide diamond-shaped achromatic bandwidth and very narrow chromatic bandwidths in the spatio-temporal Fourier domain, in parallel with humans.
\green{And this similarity is larger in the linear and shallow autoencoders (RMSE~$\sim 17\%$ of the maximum sensitivity) while it decays for deeper networks (RMSE~$\sim 22\%$ of the maximum sensitivity), again due to poor scaling of the chromatic CSFs (Fig.~\ref{Exp5_csfs}).}

\item The gain in shallow autoencoders decays with contrast and hence the contrast responses for gratings saturate with contrast. This happens both in the spatial and the spatio-temporal cases (Figs.~\ref{nonlinear_resp_color} and~\ref{nonlinearities_spatio_temp}). This resembles contrast masking in humans. However, in deeper autoencoders this consistent saturation (and hence similarity with humans) is not found.

\end{itemize}


\blue{The emergence of human-like features in the CSFs (distinct bandwidth and shape of achromatic and chromatic channels) is related to the different properties of achromatic and chromatic patterns in visual scenes.
The statistical unbalance towards achromatic patterns is known from long ago in terms of variance~\cite{Ruderman98} and more recently, it has been quantified in accurate information theoretic units~\cite{Malo20}.
The eventual problems in preserving the saturation (or poor scaling of chromatic CSFs) in deeper models, do not come from training.
Note that, according to the learning curves, all the models achieved proper convergence.
On the contrary, the problems may come from the small (statistical) relevance of chromatic textures as opposed to the achromatic textures
and the inability of deeper models to deal with this unbalance with a low-level $\varepsilon_{\textrm{LMS}}$ goal:
(too) flexible networks optimized to compensate the distortions focus (too much) on the spatial achromatic information to optimize the goal and are likely to distort chromatic information. The consequence is a negative impact on the chromatic CSFs. This does not seem to be a problem for more rigid shallower architectures and even the linear solution.}


At this low abstraction level, where the minimization of distortion in LMS is simply connected to information maximization, \green{and in the set of architectures considered,} shallow networks seem more appropriate to explain the CSFs.

\subsection{Relation to other accounts of the CSFs}
Our results revisit classical work on the statistical grounds of the CSFs~\cite{Atick92,Atick92a,Atick92b} in light of the new possibilities provided by automatic differentiation.

From the technical point of view,
a number of assumptions that had to be done in the 90's, either have been confirmed with the use of large data sets, or are not necessary with the use of more flexible models.
In particular, regarding the signal, Atick et al. assumed translation invariance, independence between color and space-time, and 2nd order relations (autocorrelation with $1/|f|^2$ decay).
Moreover, regarding the model, they restricted themselves to linear solutions similar to Wiener filters.
More recent studies with colorimetrically calibrated scenes~\cite{Gutmann14} have confirmed
the correctness of the color/space independence assumption.
However, the focus on the power spectrum and the linear solutions has proven to be too limited for denoising~\cite{Gutierrez06}. Adaptive (nonlinear) models that take into account additional features of the signal are required.
Nevertheless, the nonlinear networks considered here turn out to be roughly translation invariant~\cite{Gomez20b} as expected from their convolutional architectures and the stationary nature of the problems they face.
Other technical difference is in the formulation of the statistical goal:
Atick et al. maximized the mutual information between the clean signal and the response $I(\vect{x}_c,\vect{y})$; while here we minimized $\varepsilon_{\textrm{LMS}}$ between the clean signal and the response, $|\vect{X}_c - \vect{Y}|_2$.
These goals are exactly equivalent when the difference between clean signal and the response is Gaussian, which is not the case in general.
However, note that these goals are always related because the limit $|\vect{X}_c - \vect{Y}|_2 \rightarrow 0$ implies $I(\vect{x}_c,\vect{y}) \rightarrow \infty$.
Beyond the spatio-chromatic case, in our work we check the emergence of the CSFs with spatio-temporal signals, which was mentioned but not addressed by Atick et al. Finally, the consideration of nonlinear models allows us to show that the error-minimization goal may also lead to saturation of the contrast responses which, of course, was not possible in the linear framework of Atick et al.

As stated in the introduction,
other group has been working independently on the emergence of CSFs in ANNs~\cite{Arash21}.
Their results are restricted to the spatial CSF (no chromatic nor spatio-temporal cases) and are based on networks trained for higher-level goals such as classification. Therefore, their results from higher-level goals are complementary to ours, obtained from lower-level goals intended for the analysis of early visual stages such as the LGN.
\green{More generally, higher level goals such as classification performance may be an indirect way to impose preservation of colorfulness or a proper scaling of the chromatic CSFs. While chromatic information may have small relevance to minimize $\mathbf{\varepsilon}_{\textrm{LMS}}$ in reconstruction, it may be more crucial for recognition.}

\blue{Other works have obtained center surround sensors
by optimizing a linear+nonlinear network with a low-level infomax+energy goal~\cite{Karklin11}, or deeper nets with higher-level classification goals~\cite{LindseyICLR19}. These sensors could induce CSF-like bandwidths in the corresponding models but this aspect was not addressed in these works.}

\subsection{Individual Non-Euclidean distances from the optimization of average Euclidean distance}

An interesting consequence of our low-level result is that the Euclidean measure, $\varepsilon_{\textrm{LMS}}$, averaged over the set of natural images leads to systems that measure individual differences in non-Euclidean ways.
Note that in the systems that we trained given two input signals $\vect{x}$ and $\vect{x}+\Delta \vect{x}$, the difference between the corresponding responses is $\Delta \vect{y} \approx M \cdot \Delta\vect{x}$.
As a network should assess the difference between the two signals from $\Delta \vect{y}$, the \emph{perceived} difference for the system will depend on $M$.
Specifically~\cite{Epifanio03,Laparra10},
the perceived distance for the system will depend on the metric, $M^\top \cdot M$, and hence it will depend on $\lambda^2$ (the eigenspectrum of $M$) or, as seen here, on the $\textrm{CSF}^2$.
This metric is non-Euclidean: for instance, high-frequency distortions will be less relevant for the network than medium or low-frequency distortions. Even though here we did not check the correlation between the image distortions perceived by humans and networks, the observed CSFs in the networks are consistent with the fact that the Euclidean distance between images at the retina is not a good representation of human distortion metrics~\cite{Wang09,Laparra10,Hepburn20}.

The emergence of a non-Euclidean distance from the minimization of the average Euclidean distance over natural images is a counter intuitive consequence of the highly nonuniform distribution of natural scenes: distortions in less populated regions of the image space (e.g. in high frequency directions or in chromatic channels) have to be underrated to favour the average match to the data in highly populated or more informative regions (low frequency, achromatic patterns).

\blue{Recent work on autoencoders with low-level rate-distortion constraints on natural images has shown the emergence of non-Euclidean distances correlated with human opinion of distortion~\cite{Hepburn21}. Human opinion of distortion is known to be strongly mediated by the CSFs, but the bandwidth of this autoencoder and its eventual similarity with the CSF was not explored in that work.}
\subsection{Alternative low-level computational goals}

\green{Here we considered different low-level alternatives to the retinal signal enhancement goal proposed by Atick et al.: while our results are conclusive regarding the role of chromatic adaptation, more research is definitely required about the relative relevance of bottlenecks in shaping the CSFs.}

\green{On the one hand, given the small role of spatial information in the changes of the LMS image purely due to changes in illumination, it is not surprising that systems designed for pure chromatic adaptation have wide (all-pass) behavior in all channels (i.e. no spatial effect). As a consequence (as confirmed by our results) pure chromatic adaptation does not lead to CSFs with human-like shape.
This shape and relative bandwidths have to be related to other goals (e.g. the compensation of bio-distortion). However
training for chromatic adaptation does introduce an important human-like behavior (which may not emerge from other tasks): it leads to adaptive global scaling of the red-green or yellow-blue CSFs.
This effect is consistent with the observations done on spatio-chromatic adaptation under changes in spectral illumination~\cite{Gutmann14}: the spatial structure of the receptive fields remains almost constant but their chromatic tuning basically changes according to Von-Kries adaptation.
}

\green{On the other hand, as opposed to chromatic shifts, the spatial effect of bottlenecks is relevant. However, we only explored a small range of \emph{architectural} bottlenecks: the toy examples of Fig.~\ref{Archi}-right.
In this restricted set our results suggest that the compensation of the bio-degradation at the retina plays a stronger role in the emergence of human-like CSFs than the consideration of the bottlenecks.
However, bottlenecks in architectures C, D and U-Net favour the emergence of nontrivial frequency selectivity. This would be consistent with~\cite{LindseyICLR19}, who reported positive effects of bottlenecks in the emergence of center surround receptive fields.
Nevertheless, the specific configuration of the bottleneck that maximizes the human nature of the CSFs and the relative role of bottlenecks in the compensation of the retinal distortion are interesting matters for further research.}

\blue{More generally, other low-level goals could be considered together with the distortion, as for instance the \emph{information} or the \emph{energy} of the signal.
Architectural bottlenecks considered here or in~\cite{LindseyICLR19} indirectly constrain the energy and the entropy of the signal by reducing the dimensionality of the signal.} However one could consider
more general factors beyond the dimensionality as for instance the neural noise, the PDFs of signal and noise and the redundancy of the visual signals in the representation.
In fact, transmitted information may be modulated by changes of the representation and by the amount of noise even without changes in the dimensionality~\cite{Malo20}.

\green{In a separate study~\cite{Hepburn21} we have shown that rate-distortion bottlenecks in autoencoders induce distance measures which are correlated with subjective opinion of distortion.
The autoencoders we presented here do not include constraints on information, but the
emergence of a non-Euclidean metrics depending on $M$ (and hence on the CSFs) suggests that the distance will be correlated with human opinion in line with~\cite{Hepburn21}.}

\blue{Alternative low-level goals could include \emph{non-human} retinal degradation. Other species have different optical quality and noise in their retinas may be substantially different. This may affect the kind of computations required to extract the appropriate information from this degraded input, and hence their corresponding CSFs.}

\blue{All these issues, the specific impact of more sophisticated bottlenecks in the CSFs, which was not analyzed here nor in ~\cite{Karklin11,LindseyICLR19,Hepburn21}, the emergence of human-like image distortion measures from the enhancement of retinal signals, and the consideration of retinal degradation for other species, is a matter for future research.}

\subsection{Goal and architecture are not independent}

More important than the technical generalizations over~\cite{Atick92,Atick92a,Atick92b} is that the current freedom to explore different linear and nonlinear architectures stresses the relevance of the architectural constraints.
The conventional interpretation of the efficient coding hypothesis~\cite{Barlow1961} is the following: obtaining human-like results from certain statistical goal seems to suggest that the human visual system has been shaped by this goal.
However, it is important to realize that the results have been obtained via the optimization of certain model. In the case of Atick et al. it was a single model (the linear filter), but in our case here we tried a range of models (architectures).
Since the results for the different architectures is markedly different, the conclusion can not be about the \emph{goal}, but about specific combinations of \emph{goal-and-architecture}.
Our results are a specific illustration of the fact that the computational and the algorithmic levels of analysis of visual processing systems~\cite{Marr76,Marr82} are not independent~\cite{Poggio21}.
This dependence prevents about premature conclusions about the organization principles at the computational level if sensible architectures are not adopted.


\subsection{Beyond accuracy}

Human-like CSFs are obtained for shallow autoencoders (2-layers), or even linear networks, despite deeper architectures achieve similar or better performance in the goal. Previous literature has warned about the limitations of a single accuracy/performance measure to identify human-like behavior. Achieving similar performance on a task does not guarantee that two models actually use the same strategy~\cite{Firestone20}.
For instance, different strategies may become evident if performance degrades in different ways when changing
the experimental setting~\cite{WichmannHVEI17,GeirhosICLR19}.
Therefore, additional checks different from the optimization goal have to be done in order
to confirm the human-like behavior of a model.
Examples include verifying additional psychophysics not included in the goal~\cite{Martinez19},
or disaggregating the results checking the consistency between model and humans
in individual trials, not on averages over the data set~\cite{Geirhos20beyond}.

\blue{In this complexity/accuracy discussion, it is important to stress that our results (shallower networks better reproduce the scale of human chromatic CSFs) is in line with the results of~\cite{Gomez20b,Flachot20} which also show that shallower networks obtain more human-like colour representation.
In a similar vein, although using higher-level classification goals, \cite{KubiliusNEURIPS2019} show that lower performance networks may correlate better with human brain activity or psychophysics.}

\subsection{Final remarks}

In visual neuroscience, deep models are emerging as the new standard to reproduce the
activity of visual areas under natural scene stimulation.
On the one hand, conventional deep models driven by object recognition goals reproduce the response
from V1~\cite{Kriegeskorte15,GucluJN15}, dorsal and ventral streams~\cite{Cichy16}, and IT~\cite{Cadieu14,Yamins14}.
On the other hand, deep networks are powerful enough to fit the mappings between stimuli and measured responses~\cite{Gallant04,Antolik16,Batty17}.
These two approaches (goal-driven and measurement-driven deep models)
have been thoroughly compared in V1 and were found to be superior to
linear filter-banks and simple linear-nonlinear models~\cite{Cadena19}.
However, more recently, the same team has shown that linear-nonlinear models with general Divisive Normalization make a significant step towards the performance of state-of-the-art CNN with interpretable parameters~\cite{Cadena20}.

In our \blue{low-level} goal-driven case, the emergence of human-like CSFs for certain CNN autoencoders generalizes in different ways previous statistical explanations of the CSF based on linear models~\cite{Atick92,Atick92a,Atick92b}, \blue{and is consistent with optimizations of nonlinear encoders using alternative low-level~\cite{Karklin11,Hepburn21} or higher-level~\cite{LindseyICLR19,Arash21} goals.
However, we find a strong dependence of the CSFs on the architecture
with better results for shallower autoencoders (although they have similar or lower performance in the goal).}

\blue{This is not in contradiction with the literature cited above showing that deep networks with object recognition goals match very well higher visual areas. Note that the scope of our low-level goal is restricted to early visual stages (e.g. the retina-LGN path), and hence simpler architectures may be required there.}
%
%

\blue{Beyond this difference in abstraction-level, our results do illustrate the relevance of using appropriate architectures when checking a statistical goal.
Following the move from conventional CNNs in~\cite{Cadena19}
to more realistic divisive normalization models in~\cite{Cadena20},
we think that future goal-driven derivations of low-level visual psychophysics (e.g. pattern masking or perceptual distortion) should include more realistic architectures too, as opposed to conventional CNNs (although they may be flexible enough to fulfill the goal). Examples include divisive normalization with parametric interaction between  features~\cite{Martinez18,Martinez19} and generalizations of Wilson-Cowan interactions~\cite{SciRep20}.
Learning frameworks with rate-distortion bottlenecks are already available~\cite{Balle17,Hepburn21}, and we advocate for the study of their artificial psychophysics using realistic and interpretable architectures.}

\section{Acknowledgements}

This work was partially funded by the grants:  MICINN DPI2017-89867-C2-2-R,MICINN PID2020-118071GB-I00, and GVA Grisolía-P/2019/035 (for JM and QL), and MICINN PGC2018-099651-B-I00 (for AGV and MB).


\vspace{0.5cm}


\bibliographystyle{jovcite}


\newpage

\section{Appendix A: Implementation details}

As stated in the main text, all the models follow the
the basic toy networks studied in~\cite{Gomez18,Gomez20b}: autoencoders with convolutional layers made of 8 feature maps with kernels of spatial size $5\times5$ and sigmoids or Rectified Linear Units (ReLU) as activation functions.
As illustrated in Fig.~\ref{Archi} the last reconstruction layer, has 3 features in every case (the three color channels) so that the input and output domains are the same.
Following our purpose of using biologically plausible image representations the input to the networks and the output signals are expressed in the LMS color space (as opposed to generic RGB digital counts used in the cited references).

The spatio-temporal models follow the same spirit, in this case also including convolution in the temporal dimension: autoencoders with 3D convolutional layers made of 8 feature maps with kernels of size $5\times5\times5$ and sigmoid activation functions (we didnt explore ReLU in videos because in images we found no qualitative difference between the ReLU and sigmoid results).
As in the image case, the last (reconstruction) layer for every architecture (2, 4, 6, and 8 layers) only has 3 feature maps (the LMS channels).

Implementation and training is done in the same way as in~\cite{Gomez18,Gomez20b}:
mean squared error is used as loss function in all cases and all the models are implemented using Tensorflow~\cite{Abadi2015}.
We train our models using ADAM stochastic gradient descent~\cite{Kingma} with a batch size of \blue{32} examples, momentum of 0.9, weight decay of 0.001.
In principle, a standard early stopping criterion for convergence was used based on the number of iterations with no improvement in the validation set.
However, in order to ensure appropriate convergence we visualized the learning curves and we let the iteration continue until train and validation error reached a common plateau.
All the learning curves are explicitly shown in the Appendix~C, these show that all the CSF considered in the main text come from models with the proper convergence.

\section{Appendix B: Training stimuli and stimuli for CSF estimation}

\blue{The natural stimuli to train the networks are regular photographic images from the same dataset used in~\cite{Gomez18,Gomez20b}: the Large Scale Visual Recognition Challenge 2014 CLS-LOC validation dataset (which contains 50~$\cdot 10^3$ images), leaving 10~$\cdot 10^3$ images for validation purposes.
This dataset is a subset of the whole ImageNet dataset \cite{Russakovsky2015}.
The experiments with cartoon images were done using 25~$\cdot 10^3$ frames taken from The Pink Panther Show~\cite{Freleng63} which are in the public domain.
In every case we take $128\times128$ images and assume a sampling frequency $fs = 70$ cpd, i.e. we assume that the images subtend 3.6 deg.}

\blue{The spatio-temporal models are trained over 25~$\cdot 10^3$ patches of size $32\times32\times25$ from classical Hollywood films which are in public domain: the color movies \emph{Charade}~\cite{Donen63}
and \emph{The FBI story}~\cite{Leroy59}, and the achromatic movie \emph{The Stranger}~\cite{Welles46}. In all the video cases we assume a spatial sampling of 30 cpd and temporal sampling of 25 Hz, i.e. we assume the patches subtend 1.06 deg and last for 1 sec. These somewhat arbitrary selections of the sampling frequencies (or extent of the stimuli) have mild consequences on the quantitative evaluation of the CSFs as discussed below.}

\blue{The transform from digital counts to LMS tristimulus values was done assuming the primaries and gamma curves of a standard CRT display~\cite{Colorlab}.}

\begin{figure*}[b!]
\begin{center}
\includegraphics[width=1.0\linewidth]{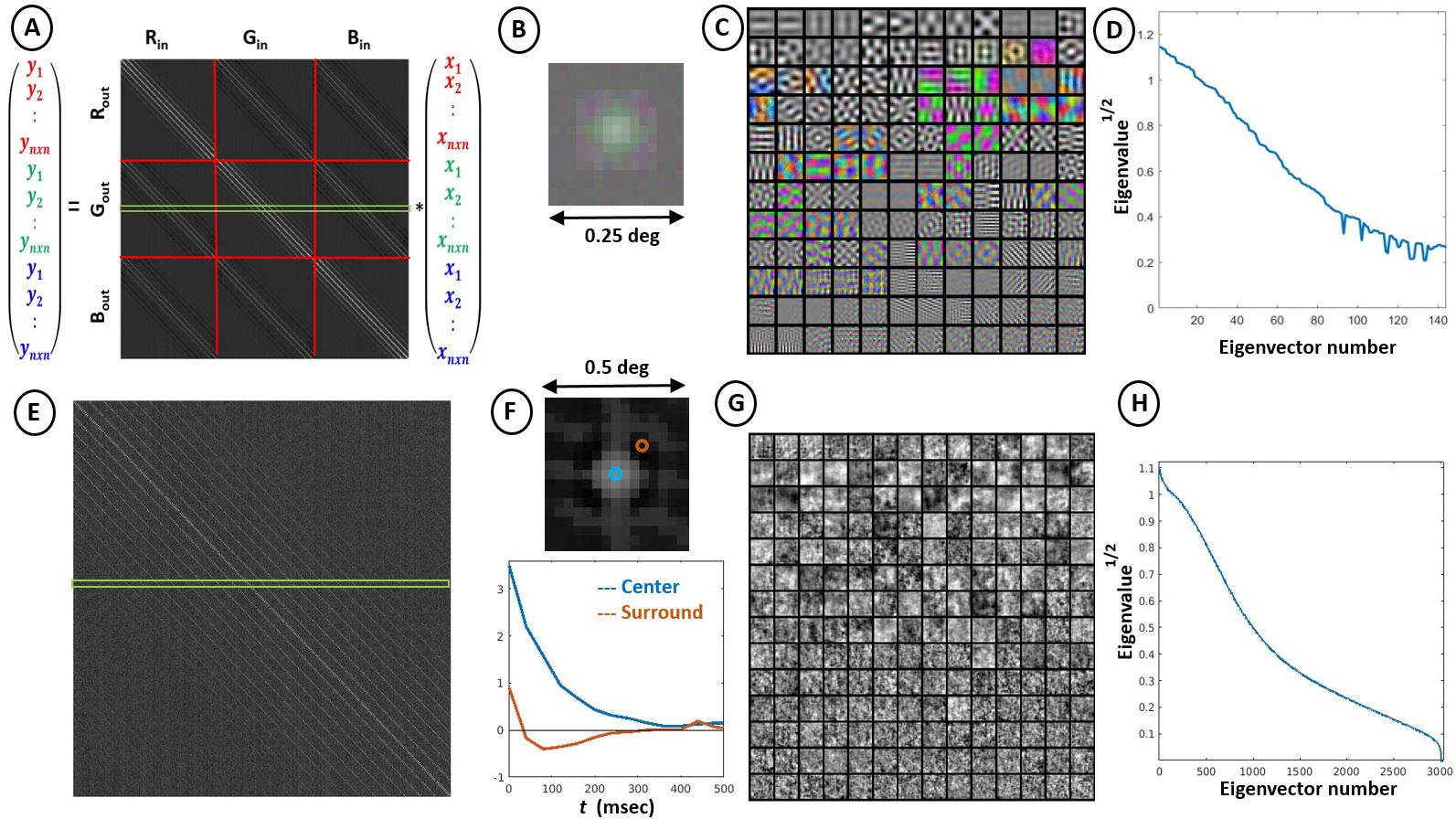}
   \caption{\textbf{Illustration of the theory: linearization and eigenanalysis.}
   \blue{The top row shows an example for spatio-chromatic networks, and the bottom row one for spatio-temporal networks (in both cases with a 2-layer architecture with sigmoid activation functions).
   The matrices $M$ obtained from Eq.~\ref{global_linear_approx} are represented in panels A and E. Panel A shows that in this matrix notation the input and output have been vectorized. As a result, the receptive fields (rows of the matrix $M$) have to be de-vectorized and reshaped as a 2D image (as in panel B), or as an image with certain temporal evolution (as in panel F). Panels C and G show the eigenfunctions of the corresponding matrices $M$, where, in panel G the temporal variations are obviously missing. Finally, panels D and H show the eigenvalues of the corresponding eigenfunctions.}}
   \label{Linearization}
\end{center}
\end{figure*}

\begin{figure*}[t]
\begin{center}
\includegraphics[width=0.8\linewidth]{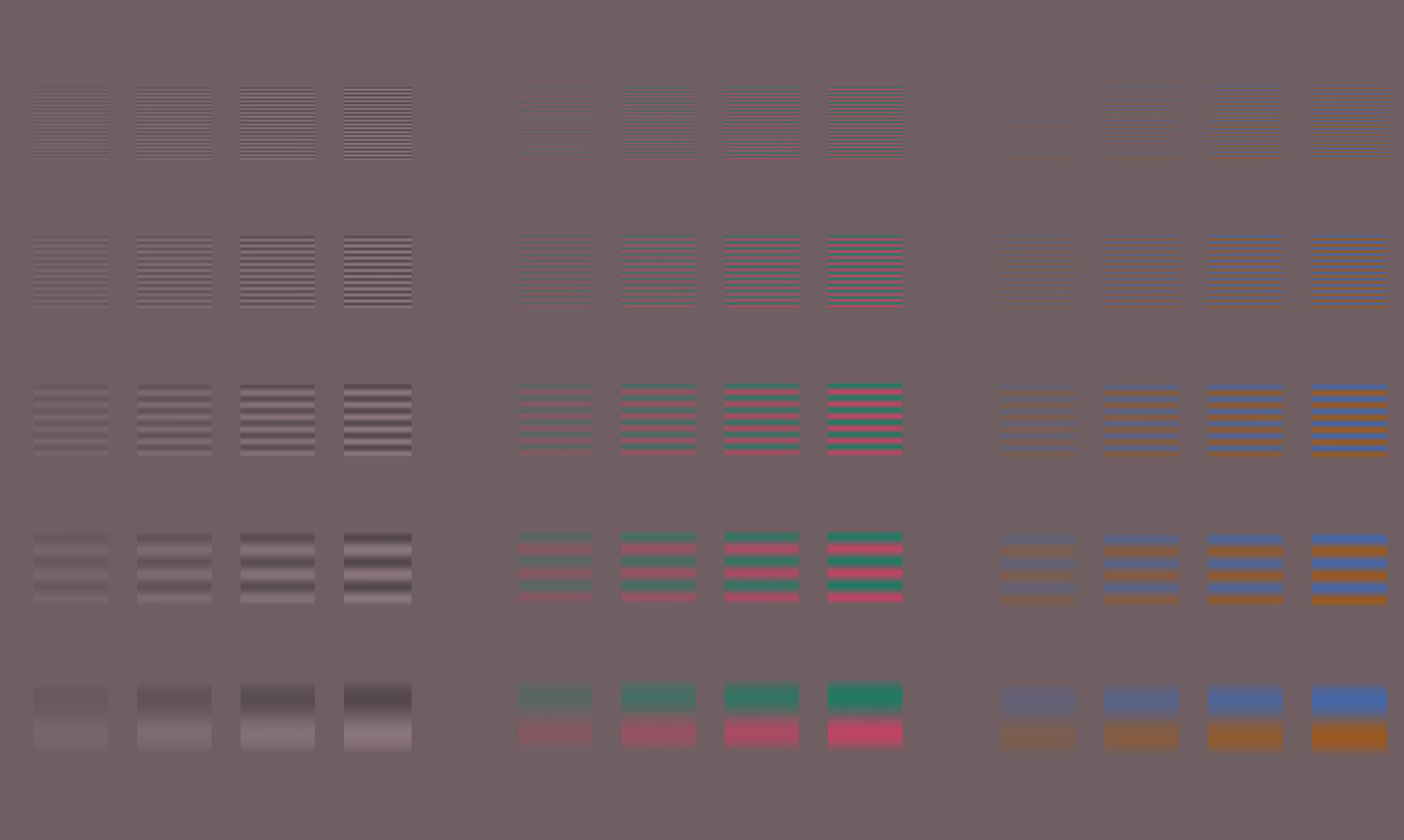}
   \caption{Representative spatio-chromatic stimuli to feed the 2D networks. The 3D networks were probed with equivalent gratings mooving at different speeds (see text).}
   \label{gratings}
\end{center}
\end{figure*}

\blue{Regarding the stimuli for the estimation of the CSFs according to Eq.~\ref{csf1}, our $\vect{b}^f$ are gratings in the classical opponent space of~\cite{Jameson59}. Figure~\ref{gratings} shows a representative subset of the gratings used to feed the networks for the estimation of the spatio-chromatic CSFs.}
\blue{The justification of the use of these waves to probe the autoencoders follows the eigenanalysis of the linearized networks introduced in~\cite{Gomez20b}:}
\green{the eigenfunctions of the matrices in Eq.~\ref{global_linear_approx}       
are oscillating functions in space with chromatic variations in luminance and opponent red-green and yellow-blue directions. Consistently with~\cite{Gomez20b} the corresponding spatio-temporal oscillations of increasing frequency for decreasing eigenvalue are obtained when the considered Jacobian corresponds to spatio-temporal autoencoders. See the illustration in Fig.~\ref{Linearization}.}
The top row just recalls the results reported in~\cite{Gomez20b} for shallow 2D spatio-chromatic networks, and the bottom row shows the new, but equivalent, results for the 3D spatio-temporal counterpart trained in achromatic movies.

Following the theory introduced in Section 2, the matrices $M$ that result from Eq.~\ref{global_linear_approx} (panels A and E) have a clear convolutional structure: rows with the same shape up to a shift in the spatial or the spatio-temporal domain. Note that in this matrix notation the input and output have been vectorized, as shown in panel A.
This structure in $M$ reveals shift invariance in the receptive fields.
Representative receptive fields are shown in panels B and F, where the row highlighted in green in the matrix has been reshaped to be represented back into the spatial or into the spatio-temporal domain.
All the linearized networks present these kind of equivalent center-surround receptive fields (a red-green example in panel B,
and an on-center/off-surround with distinct temporal responses in different spatial locations in panel F).
These receptive fields are replicated with the corresponding shift along the rows of the matrices $M$.
\blue{It is important to note that the extent of these equivalent receptive fields goes \emph{beyond} the spatial/temporal extent of the kernels used in the CNNs: they display non-zero oscillations beyond the spatial $5\times5$-pixel region, and the temporal 5-frames region.}
As anticipated in Section 2, following~\cite{Clarke81}, convolutional-like matrices lead to extended oscillatory eigenfunctions as illustrated in panels C and G.
The spatio-chromatic functions in panel C are similar to Cosine basis and interestingly, as noted in~\cite{Gomez20b}, the basis display oscillations in human-like red-green and blue-yellow opponent directions. See the original report for further analysis of the chromatic directions and color matching functions.
The spatio-temporal functions in panel G include extended textures (both in space and time), with flicker and motion, but they cannot be identified with Fourier or Cosine waves. However, it is clear that they display characteristic spatio-temporal frequencies (medium to low spatial frequencies at the top versus high spatio-temporal frequencies at the bottom).
The functions in panels C and G are ordered according to the eigenvalues (in panels D and H). This means that the systems favour medium and low spatio-temporal frequencies.
The structure of the matrices found and the order in the oscillating eigenfunctions reveal the suitability of Fourier-like basis and the higher relevance of medium and low frequencies.


The full spatio-chromatic set included gratings of \blue{60 spatial frequencies linearly spaced in the range [0.5, 35]~cpd (the Nyquist region assuming $f_s = 70 cpd$)}, and 9 contrasts linearly spaced in the range [0.07,0.6]. The average color was the white of the color system with 30 $cd/m^2$.


The stimuli for the computation of the spatio-temporal CSFs were moving sinusoids with 16 spatial frequencies in the range [0,15]~cpd, 10 temporal frequencies in the range [0,10] Hz, and 9 contrasts in the range [0.07,0.6]. The average color and luminance were the same as in the image case.

\blue{The lower limit of the explored contrast range comes from the limitation due to noise discussed in Section~2.
The upper contrast limit and the average luminance were selected to ensure that corrupted signals are reproducible in regular displays (which is the range of the scenes used in the training).}

\section{Appendix C: Convergence and Performance of the models}

\blue{In order to guarantee that the presented CSF results do not come from eventual training artifacts, this Appendix illustrates the proper convergence and proper performance of all the considered CNNs in all the goal/architecture scenarios.
For each considered case in the Experiments, we show the learning curves and explicit examples of the responses (reconstructed signals in test).
Finally, as an illustrative example, we also show one extra case (for video -Experiment 6-) where convergence was not complete in one of the networks and the consequences in the reconstructions and in the CSFs.}

\textbf{Experiment 1: distortion compensation from a range of CNN architectures}. \blue{Figure~\ref{Exp1_converg_perform} shows the learning curves of all the CNN models used in Experiment 1.
Throughout the Appendix the gray/black curves refer to the $\varepsilon_{\textrm{LMS}}$ distortion of the retinal signal in the LMS color space of \cite{Stockman00}, which is constant along the learning. The cyan/blue curves show the evolution of the error in the response of the networks. The light color curve describes the error over image batches in the \emph{training} phase while the dark color describes the same error in the \emph{validation} set. The error of the response (solution) significantly drops below the error of the input signal (problem) thus indicating that the network is actually achieving the functional goal it has been designed for.
The plateau achieved by the blue curves (not only in training but, more significantly in validation) implies that a steady convergence was achieved and the resulting model is ready to be tested.
Consistency between the train and validation sets is apparent from the parallel behavior of the light and dark curves. Performance tables in the main text (Table~1 for Exp.~1) and performance in the visual examples shown here (Fig.~\ref{Exp1_converg_perform} for Exp.~1) refer to an independent \emph{test} set not used in the learning (training/validation) phase.
In all CNNs used in Exp.~1, the training has been done in a representative set since the errors in the independent test phase (Table~1 and Fig.~\ref{Exp1_converg_perform}) are consistent with the asymptotic behavior of the learning curves shown in Fig.~\ref{Exp1_converg_perform}.} \blue{Figure~\ref{Exp1_visual_perform_lin} shows visual examples of the performance of the linearized versions of the nonlinear models in Exp.~1. It is interesting to note that the optimal linear solution (computed from the train set) has worse behavior than the linearized versions of the networks (as also seen in Table~1)}.

\afterpage{
\begin{landscape}
\vspace{5cm}
\begin{figure}[!h]
\vspace{-0.0cm}
   \centering
\includegraphics[width=0.85\textwidth]{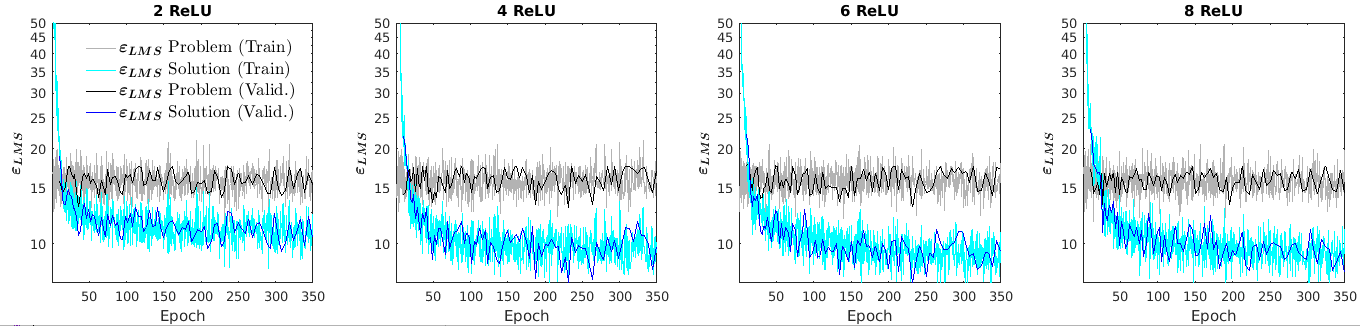}\\
\includegraphics[width=0.85\textwidth]{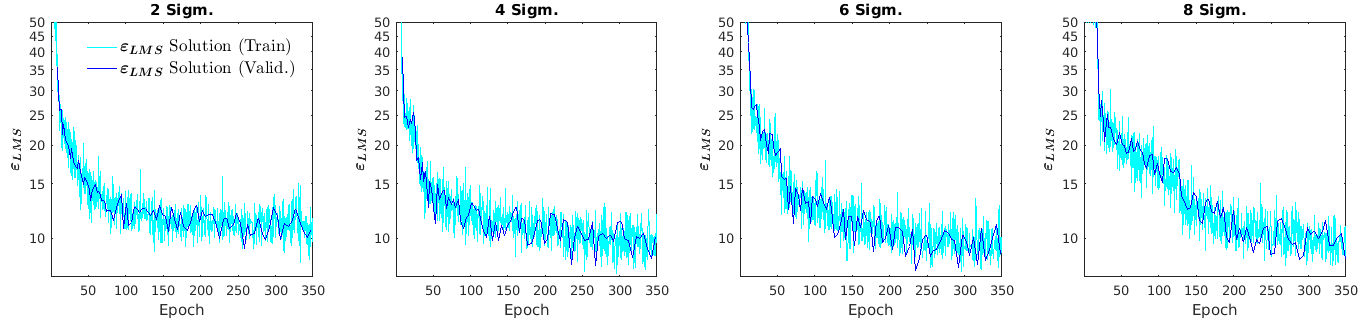}\\[0.5cm]
\hspace{-0cm}\includegraphics[width=1.22\textwidth]{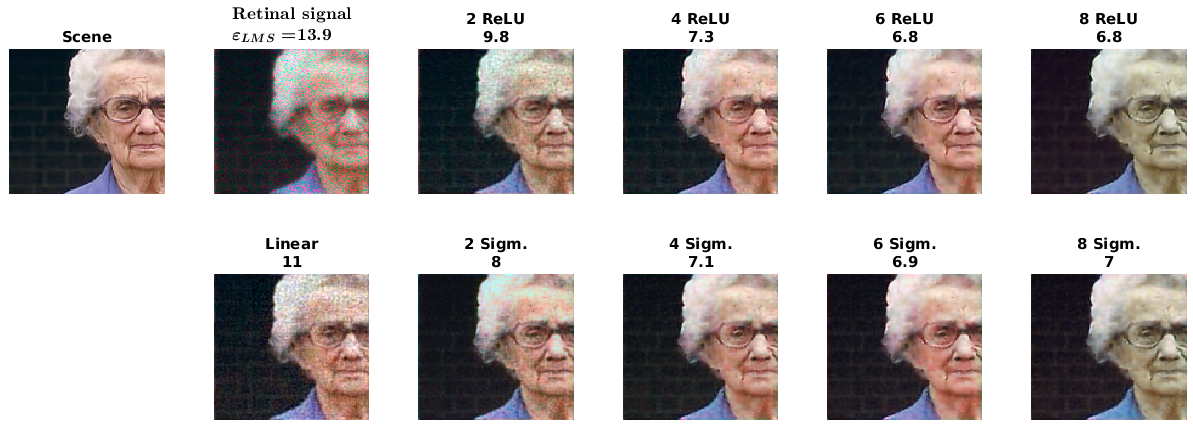}\\
   \caption{\textbf{Experiment 1: Convergence and trained models} Learning curves (training/validation) and examples of visual performance (test) of all the models trained in Experiment 1. \green{The distortion $\varepsilon_{\textrm{LMS}}$-Problem refers to the original degradation of the images (previous to the application of the net). This distortion describes how difficult the compensation problem is. The distortion $\varepsilon_{\textrm{LMS}}$-Solution refers to the degradation remaining in the signal after the application of the net. It describes how close the output is to the ideal result.}
    Performance numbers have been truncated to the significance of the standard deviation in the test set.}
   \label{Exp1_converg_perform}
\end{figure}
\end{landscape}
}

\afterpage{
\begin{landscape}
\vspace{5cm}
\begin{figure}[!h]
\vspace{4cm}
   \centering
\hspace{-1cm}\includegraphics[width=1.22\textwidth]{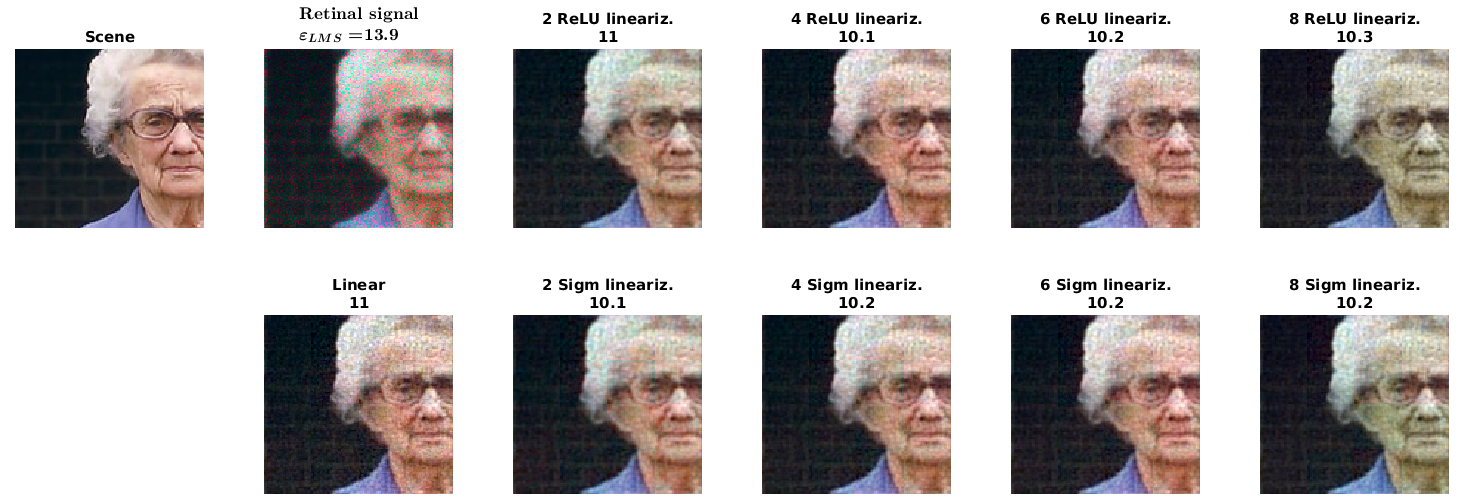}\\
   \caption{\textbf{Experiment 1: linearized models.} Examples of visual performance (in test) for the linearized CNNs of Experiment 1.}
   \label{Exp1_visual_perform_lin}
\end{figure}
\end{landscape}
}



\textbf{Experiment 2: architecture trained on a range of distortion levels}
\blue{Figure~\ref{Exp2_converg} shows the learning curves of the model trained in Experiment 2 (2-layers ReLU) for different levels of retinal degradation (noise/blur). Note that the specific cases where the iteration was stopped due to the activation of the early stopping criterion (top-left and bottom-center), the convergence plateau was already reached. Figure~\ref{Exp2_performance} shows examples of the performance in test of the model considered in every training scenario considered in Experiment 2.}

\afterpage{
\begin{figure}[b]
\begin{center}
\includegraphics[width=0.75\linewidth, height=38em]{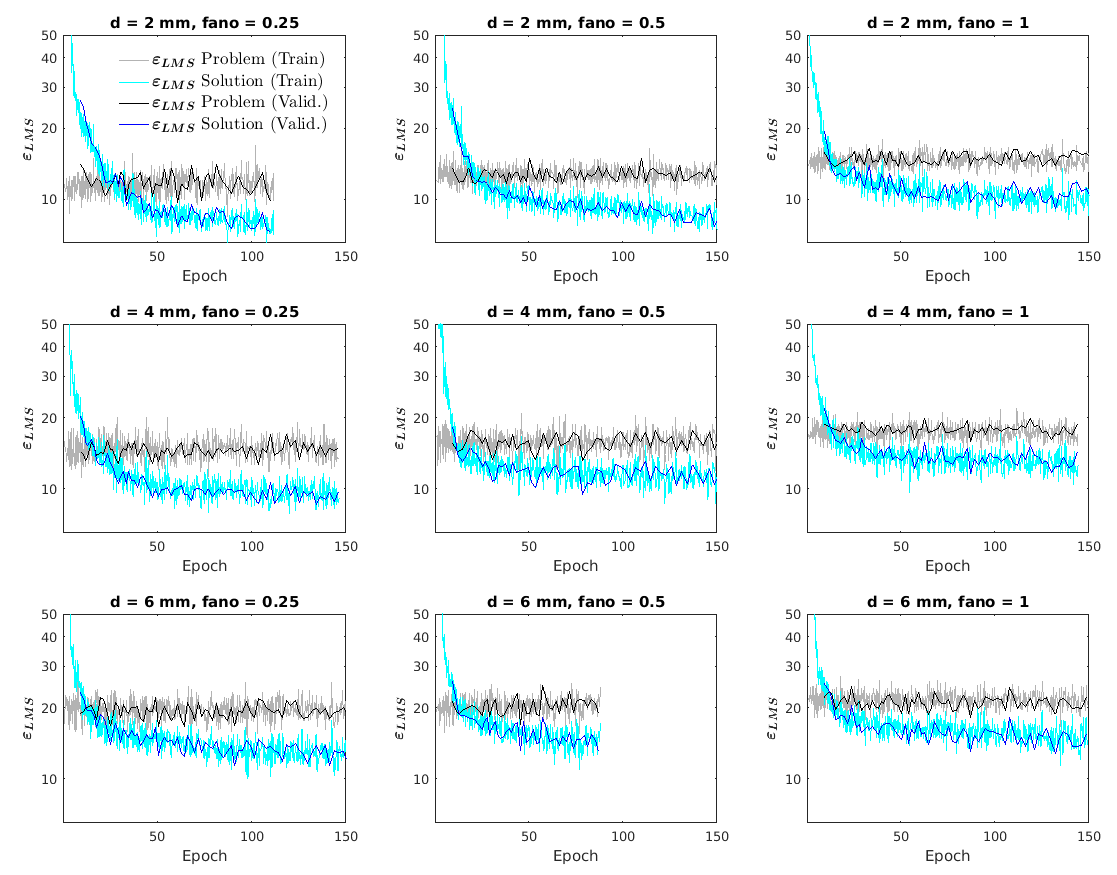}\\[1cm]
\includegraphics[width=0.75\linewidth, height=9em]{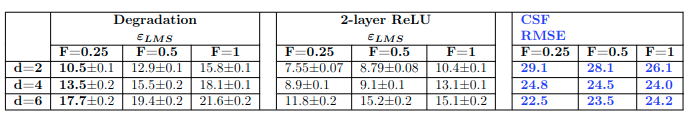}\\
   \vspace{-0.2cm}
   \caption{\textbf{Experiment 2: Convergence and performance} Learning curves (training/validation) and numerical performance (in test and in the reproduction of the CSFs) of the CNN model trained in all conditions considered in Experiment 2.}
   \label{Exp2_converg}
\end{center}
\end{figure}

\begin{figure}[h]
\begin{center}
\begin{centering}
\hspace{-0cm}\includegraphics[width=0.8\linewidth, height=38em]{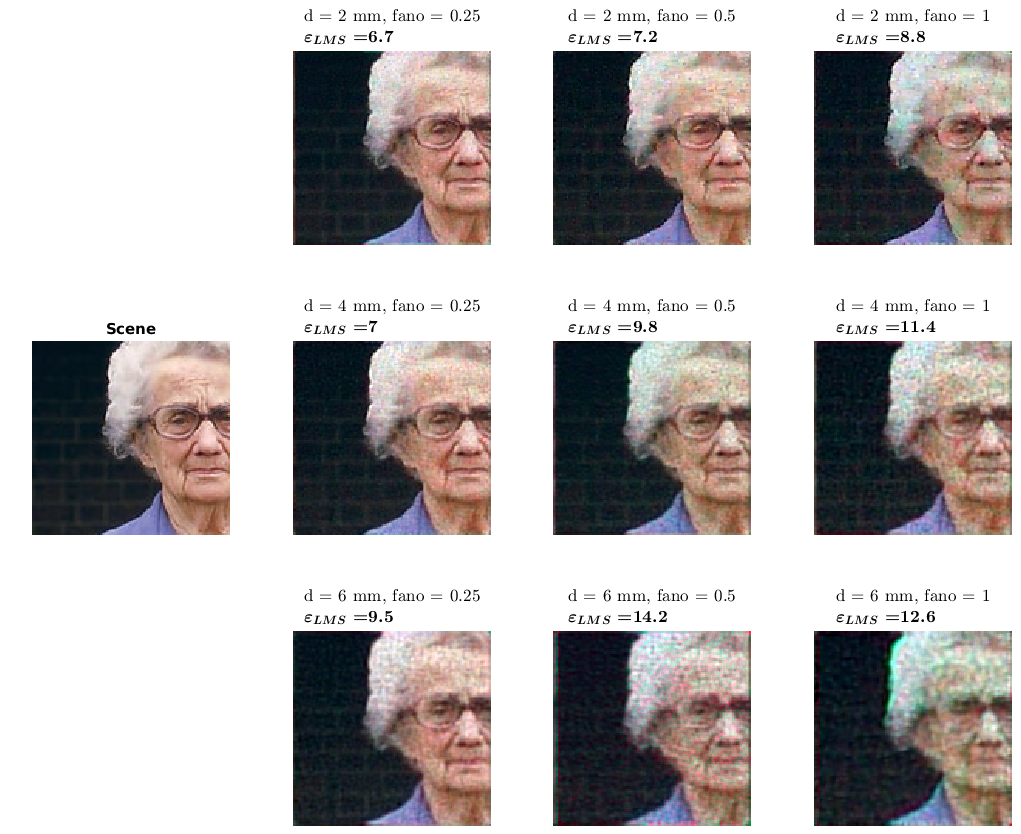}
\end{centering}
\vspace{-0.2cm}
\caption{\textbf{Experiment 2: visual performance.} Examples of reconstruction (in test) for the CNN in all the degradation scenarios considered in Experiment 2.}
   \label{Exp2_performance}
\end{center}
\end{figure}
}

\textbf{Experiment 3: chromatic adaptation versus distortion compensation} \blue{Figure~\ref{Exp3_converg_performance} (top) demonstrates that the model trained for the five computational goals considered in Experiment 3 actually achieves the goals and has proper convergence. Figure~\ref{Exp3_converg_performance} (bottom) shows visual examples of the performance.}


\afterpage{
\begin{landscape}
\vspace{5cm}
\begin{figure}[!h]
\vspace{2cm}
   \centering
\hspace{0cm}\includegraphics[width=1.22\textwidth, height=16em]{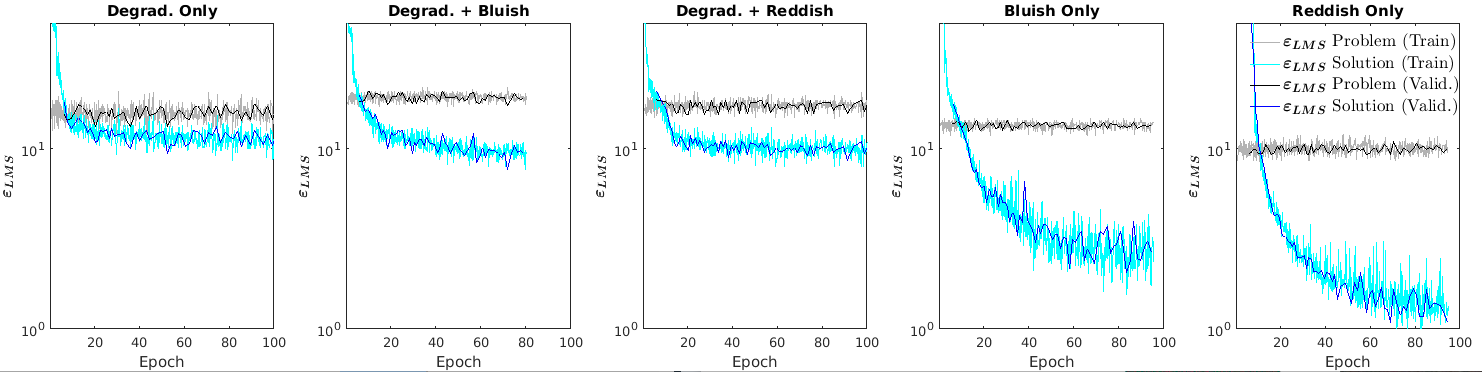}\\[0.5cm]
\hspace{0cm}\includegraphics[width=1.22\textwidth]{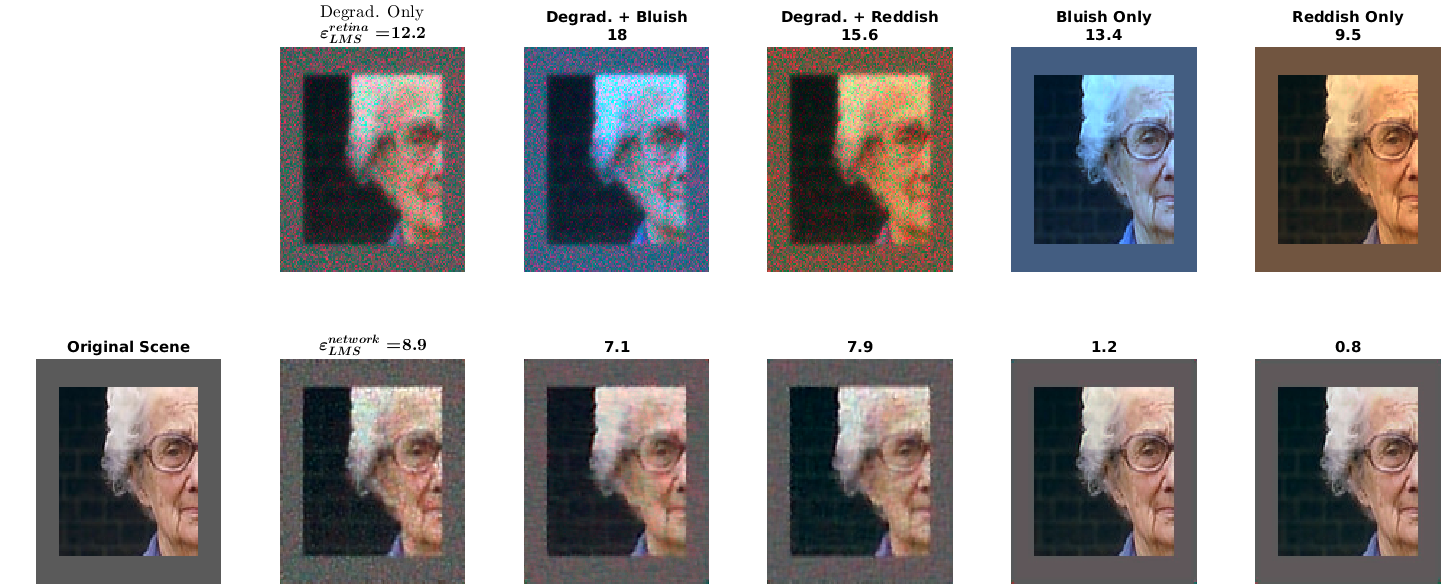}\\
   \caption{\textbf{Experiment 3: Convergence and Visual Performance:} Top panel, learning curves (train/validation) for the considered architecture in the different goals. Bottom panel: Visual example (test) for the CNNs in Experiment 3 (natural images).}
   \label{Exp3_converg_performance}
\end{figure}
\end{landscape}
}

\textbf{Experiment 4: robustness under change of signal statistics}
Figure~\ref{Exp4_converg_performance} (top) demonstrates that the model trained for the five computational goals considered in Experiment 4 actually achieves the goals and has proper convergence. Figure~\ref{Exp4_converg_performance} (bottom) shows visual examples of the performance.



\afterpage{
\begin{landscape}
\vspace{5cm}
\begin{figure}[!h]
\vspace{2cm}
   \centering
\hspace{0cm}\includegraphics[width=1.22\textwidth]{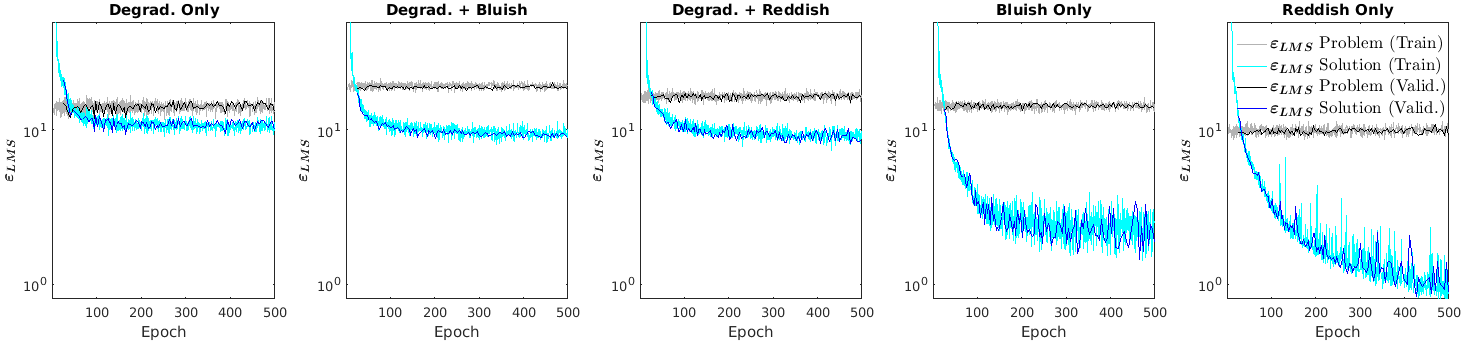}\\[0.5cm]
\hspace{0cm}\includegraphics[height=7.7cm,width=1.22\textwidth]{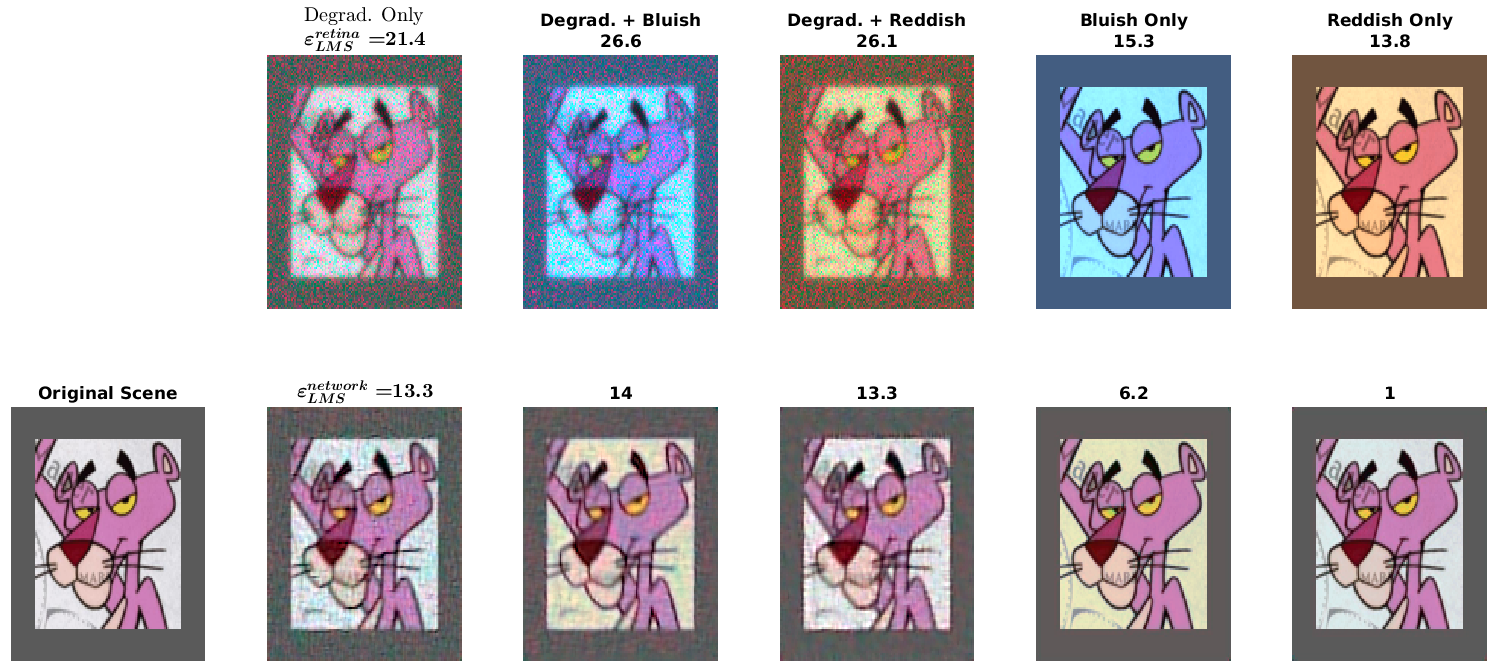}\\
   \caption{\textbf{Experiment 4: Convergence and Visual Performance:} Top panel, learning curves (train/validation) for the considered architecture in the different goals. Bottom panel: Visual example (test) for the CNNs in Experiment 4 (cartoon images). Original image from \emph{The Pink Panther Show}~\protect\cite{Freleng63} courtesy of the MGM. Similar images~\protect\cite{Malo22} lead to equivalent performance in the networks.}
   \label{Exp4_converg_performance}
\end{figure}
\end{landscape}
}

\textbf{\green{Experiment 5: CSFs from bottleneck compensation and bio-distortion compensation}}
\green{Figure~\ref{Exp5A_converg} demonstrates that the models trained for the goals considered in Experiment 5 have proper convergence and actually achieve the goals within the constraints imposed by the bottlenecks of progressive severity. Figure~\ref{Exp5A_visual_perform} shows visual examples of the performance.}
\green{The reconstruction error $\varepsilon_{\textrm{LMS}}$ behaves quite intuitively
(see Fig.~\ref{Exp5A_converg} here and  Table~\ref{Exp5A_table} in the main text):
On the one hand, in the cases that involve bio-distortion all the architectures do reduce the original value of $\varepsilon_{\textrm{LMS}}$ except the architecture $G$, which has a single feature in its bottleneck.
On the other hand, the pure reconstructions cases introduce negligible distortion $\varepsilon_{\textrm{LMS}}$ when the inner representation
does not restrict spatial resolution nor number of features (the no-bottleneck cases A and B). And, as expected, more severe bottlenecks imply higher $\varepsilon_{\textrm{LMS}}$.
}

\afterpage{
\begin{landscape}
\vspace{5cm}
\begin{figure}[!h]
\vspace{4cm}
   \centering
\hspace{-1cm}\includegraphics[width=1.22\textwidth]{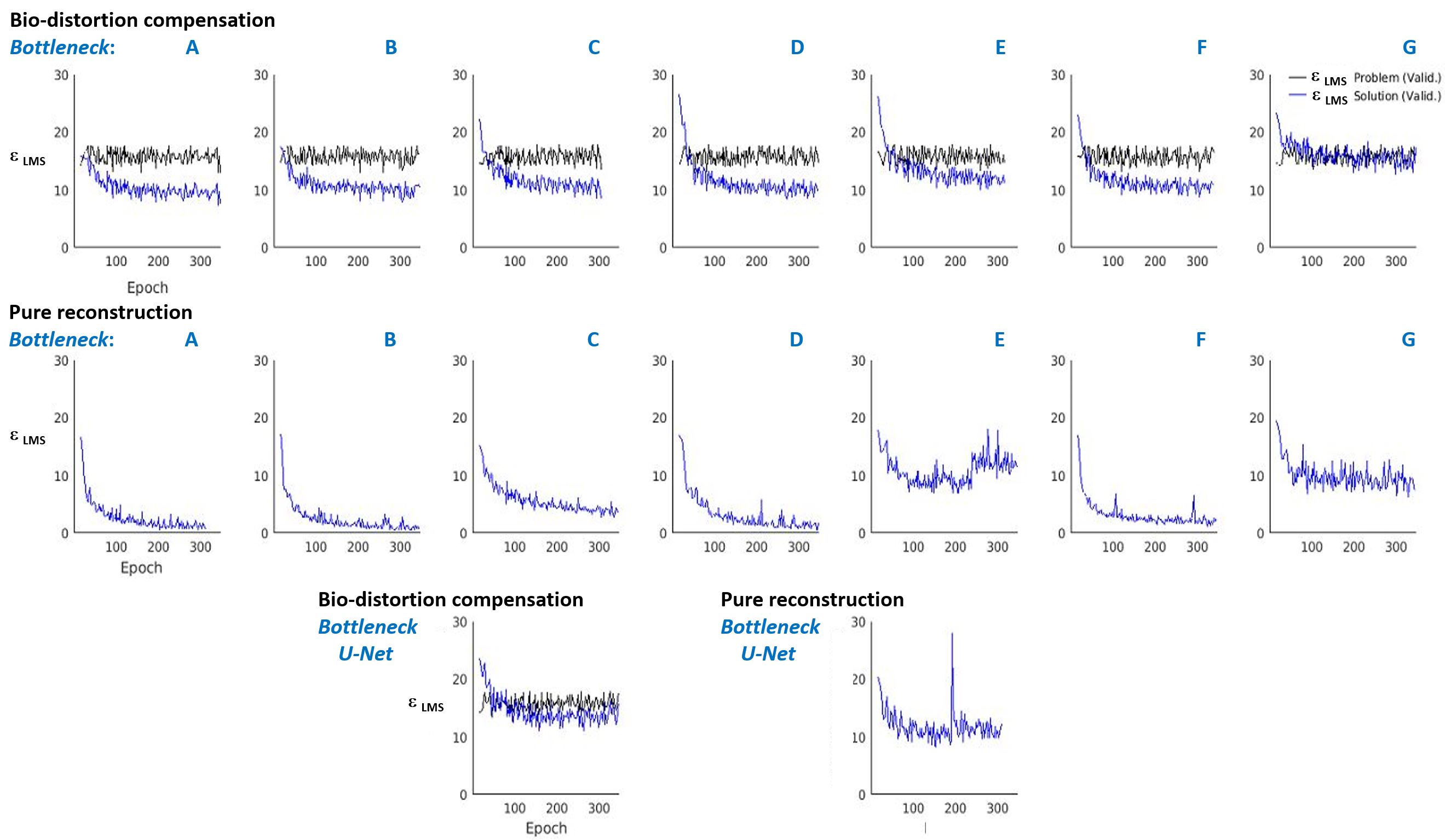}\\
   \caption{\textbf{Experiment 5: Bottleneck compensation and bio-distortion compensation.} Learning curves (train/validation) for the considered architectures in reconstruction with compensation of bio-distortion (top row and bottom row, left) and pure recosntruction (middle row and bottom row, right). The cases including bio-distotion show the original $\varepsilon_{\textrm{LMS}}$ of the problem.
   See Fig.~\ref{Archi} for the structure of the architectures referred by the letters in blue.}
   \label{Exp5A_converg}
\end{figure}
\end{landscape}
}

\afterpage{
\begin{landscape}
\vspace{5cm}
\begin{figure}[!h]
\vspace{4cm}
   \centering
\hspace{-1cm}\includegraphics[width=1.22\textwidth]{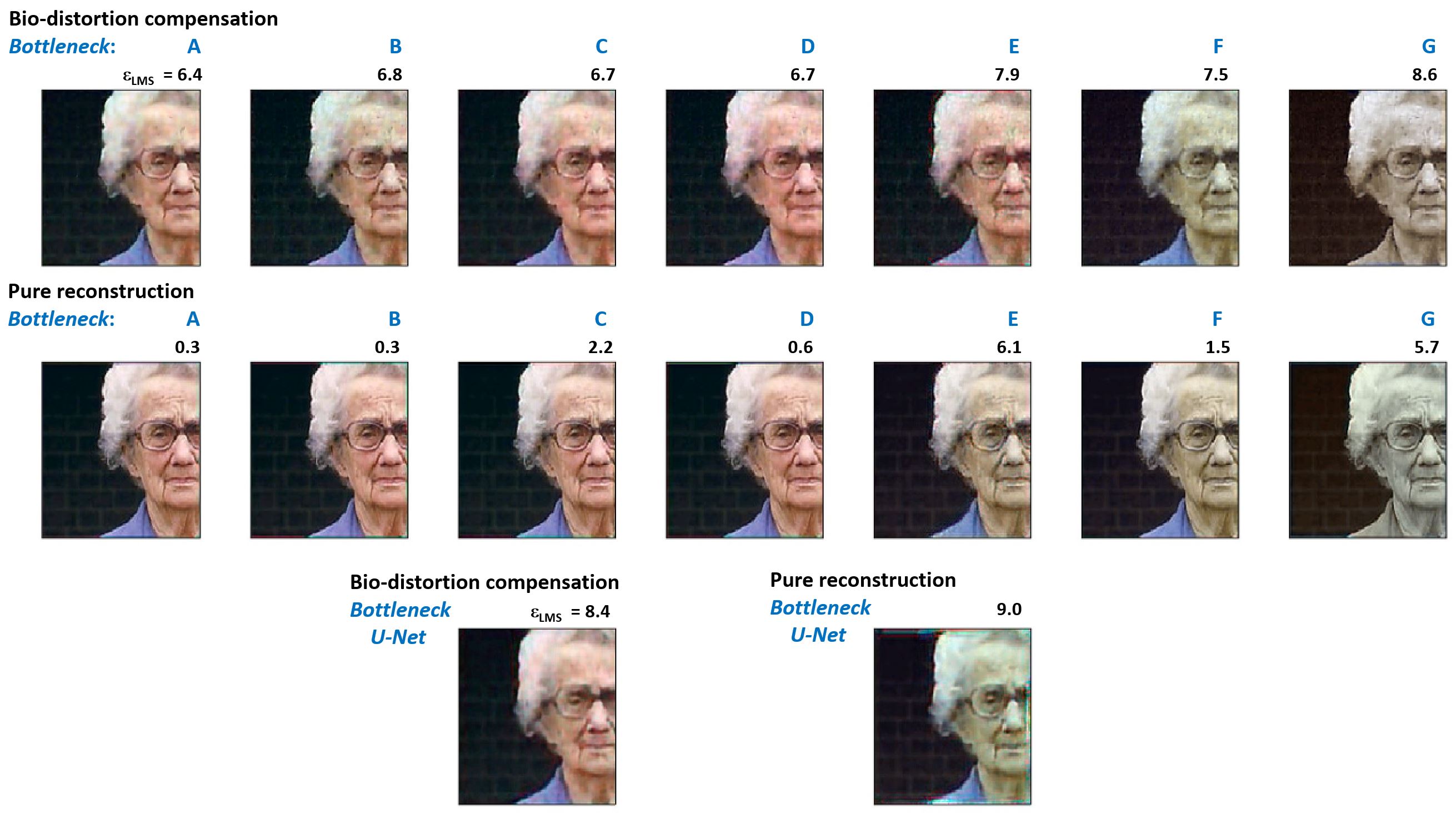}\\
   \caption{\textbf{Experiment 5: Bottleneck compensation and bio-distortion compensation.} Examples of visual performance (in test) for the CNNs of Experiment 5. See Fig.~\ref{Archi} for the structure of the architectures referred by the letters in blue.}
   \label{Exp5A_visual_perform}
\end{figure}
\end{landscape}
}

\textbf{Experiment 6: Spatio-temporal-temporal CSFs}
\blue{The main text includes the CSF results from a range of models trained in Charade, 1963. Figure~\ref{Exp5_converg_perform} shows the regular evidences on convergence (top) and visual performance in test (bottom) shown for the other Experiments.}
\blue{In this Appendix we also include a replication of Experiment 6 trained on a movie with higher spatial resolution (The FBI story, 1959).
We give the corresponding learning curves and reconstructions (Fig.~\ref{Exp5_converg_performance2}) and CSFs (Fig.~\ref{Exp5_csfs2}). This is interesting for two reasons: (1) it confirms the superiority of shallower nets even for different resolution, and (2) it shows an example of failure in convergence (see that the 8-layer model in Fig.~\ref{Exp5_converg_performance2} got stuck in a local minimum (with poor performance) and this has consequences in the complete loss of chromatic information. Also interesting is the fact (consistent in the other image/video examples in the main text), that models with 4-,6-layers (which converged as well as the 2-layer model), substantially over attenuate the red-green channel with the corresponding yellowish-bluish look of the reconstruction and the corresponding impact on the relative scaling of channels in the CSFs, which is not the case for the linear and the 2-layer solutions.}


\afterpage{
\begin{landscape}
\vspace{5cm}
\begin{figure}[!h]
\vspace{2cm}
   \centering
\hspace{0cm}\includegraphics[width=1.22\textwidth]{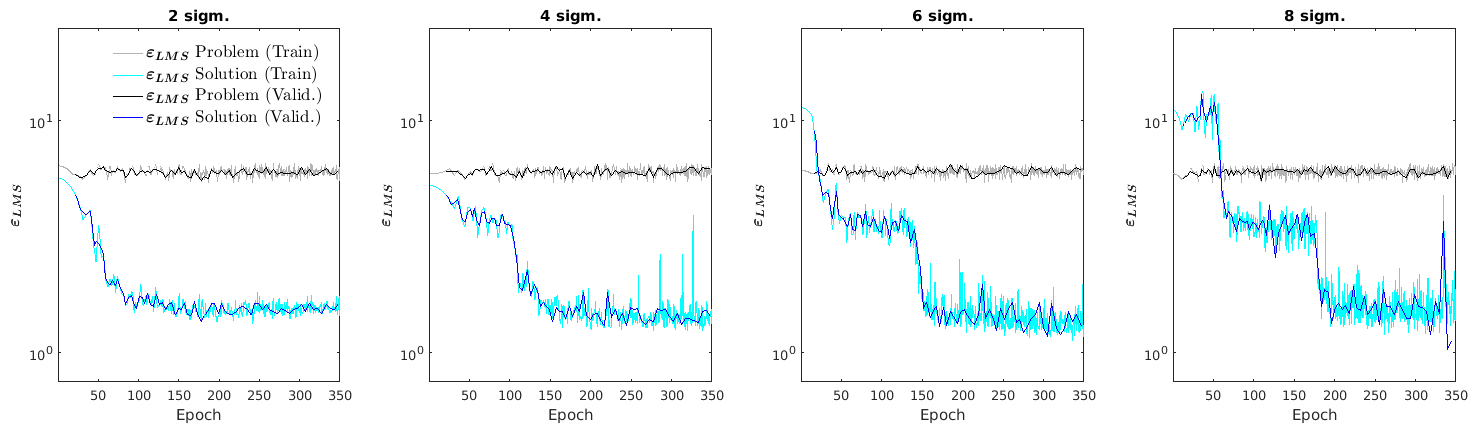}\\[0.5cm]
\hspace{0cm}\includegraphics[width=1.22\textwidth]{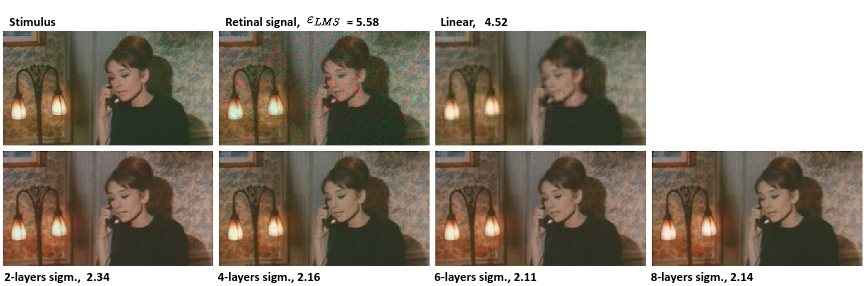}\\
   \caption{\textbf{Experiment 6 (CSFs in main text): Convergence and visual performance.}
   Top panel, learning curves for the considered architectures.
   Visual example (test) for the linear solution and the CNNs in Experiment 6 (Charade, low resolution movie). The original frame comes from the film \emph{Charade}~\protect\cite{Donen63} which is in the public domain.}
   \label{Exp5_converg_perform}
\end{figure}
\end{landscape}
}


\afterpage{
\begin{landscape}
\vspace{5cm}
\begin{figure}[!h]
\vspace{-0cm}
   \centering
\hspace{0cm}\includegraphics[width=1.22\textwidth,height=5.5cm]{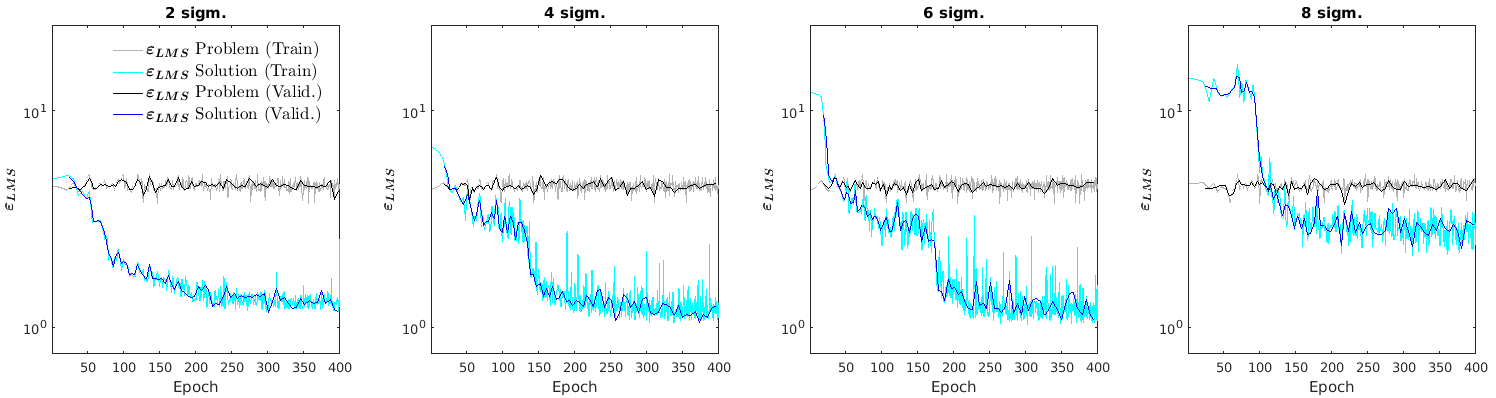}\\[-0.2cm]
\hspace{0cm}\includegraphics[width=1.22\textwidth,height=12cm]{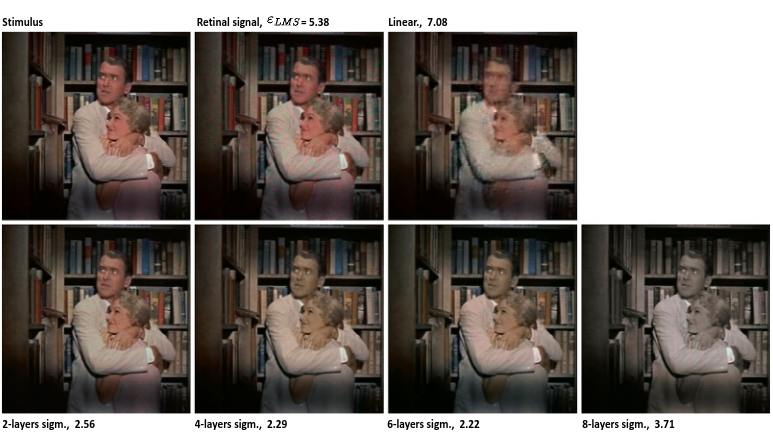}\\[-0.2cm]
   \caption{\textbf{Extended Experiment 6 (not shown in the main text): Convergence and visual performance.}
   Top panel, learning curves for the considered architectures. All converge except 8-layers.
   Visual example (test) for the linear solution and the CNNs in Experiment 6 (The FBI story, high resolution movie).
   The original frame comes from the film \emph{The FBI story}~\protect \cite{Leroy59}.
   }
   \label{Exp5_converg_performance2}
\end{figure}
\end{landscape}
}

\afterpage{
\begin{figure*}[t!]
\begin{center}
\begin{centering}
\hspace{-0cm}\includegraphics[width=0.65\linewidth]{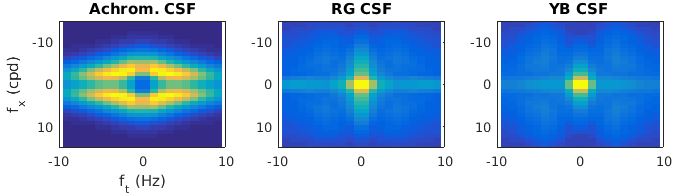}\\
\hspace{-0cm}\includegraphics[width=0.65\linewidth]{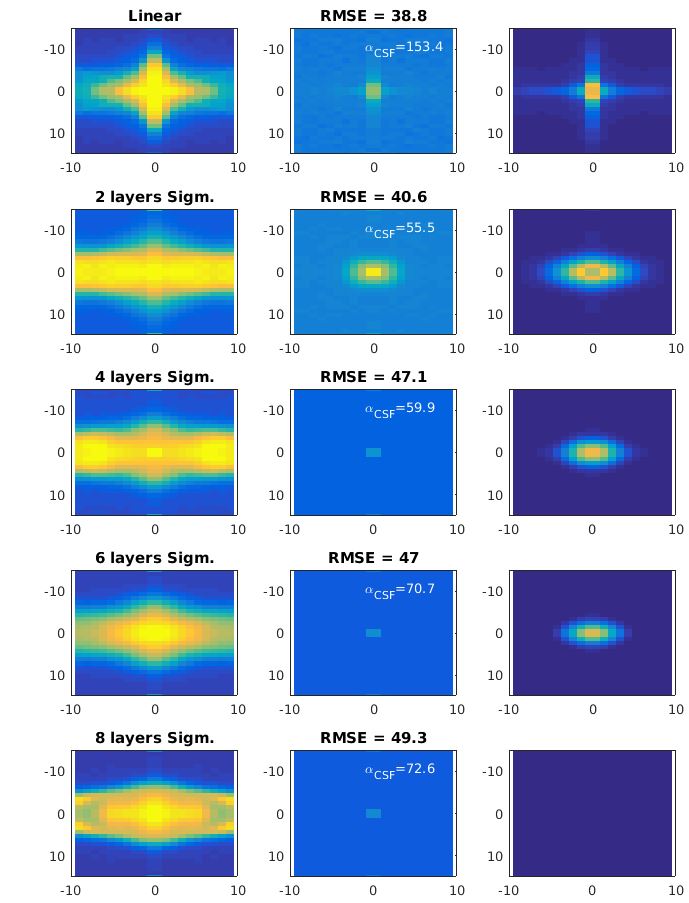}\\
\end{centering}
   \caption{\textbf{Extended Experiment 6 (CSFs of high. resolut. movie not shown in the main text)}.
   Note that in this case (see Fig.~\ref{Exp5_converg_performance2}) all models converged except the 8-layer architecture.}
   \label{Exp5_csfs2}
\end{center}
\end{figure*}
}

\end{document}